\documentclass[onecolumn, draftcls, 12pt]{IEEEtran}

\usepackage{amsfonts}
\usepackage{mathrsfs}
\usepackage{amssymb,amsmath}
\usepackage[noblocks]{authblk}
\usepackage[compress]{cite}
\usepackage{bm}
\usepackage{algorithm}
\usepackage{algorithmic}
\usepackage{amsmath,amssymb,epsfig,graphics,subfigure}
\usepackage{theorem}
\usepackage{array,color}
\usepackage[compress]{cite}
\usepackage{algorithm}
\usepackage{algorithmic}

\newtheorem{conclusion}{Conclusion}

\theoremheaderfont{\normalfont\bfseries}
\usepackage[normalem]{ulem}

\begin{document}

\title{On Weighted MSE Model for MIMO Transceiver Optimization}

\author{Chengwen Xing, Yindi Jing, and Yiqing Zhou \thanks{C. Xing is with School of Information and Electronics, Beijing Institute of technology, Beijing 100081,
China (e-mail: chengwenxing@ieee.org).}  \thanks{Y. Jing is with the Department of Electrical and Computer Engineering,
University of Alberta, Edmonton, AB T6G 2V4, Canada (e-mail: yindi@ualberta.ca). }\thanks{Y. Zhou is with the Wireless Research Center, Institute of Computing
Technology, Chinese Academy of Sciences, Beijing 100190, China (e-mail:
zhouyiqing@ict.ac.cn).} 
} \maketitle

\begin{abstract} 
Mean-squared-error (MSE) is one of the most widely used performance metrics for the designs and analysis of multi-input-multiple-output (MIMO) communications. Weighted MSE minimization, a more general formulation of MSE minimization, plays an important role in MIMO transceiver optimization. While this topic has a  long history and has been extensively studied, existing treatments on the methods in solving the weighted MSE optimization are more or less sporadic and non-systematic. In this paper, we firstly review the two major methodologies, Lagrange multiplier method and majorization theory based method, and their common procedures in solving the weighted MSE minimization. Then some problems and limitations of the methods that were usually neglected or glossed over in existing literature are provided. These problems are fundamental and of critical importance for the corresponding MIMO transceiver optimizations. In addition, a new extended matrix-field weighted MSE model is proposed. Its solutions and applications are discussed in details. Compared with existing models, this new model has wider applications, e.g., nonlinear MIMO transceiver designs and capacity-maximization transceiver designs for general MIMO networks.
\end{abstract}


\section{Introduction}
Multiple-input-multiple-output (MIMO) technology is a milestone in both wireless academia and wireless industry. By exploiting the extra spatial resources provided by the multiple antennas at the transmitter and/or receiver, MIMO technology can greatly enhance the reliability and spectral efficiency of wireless communications \cite{Larsson03,Tse05,Bolcskei06}. These two types of gains are referred to as the diversity gain and spatial multiplexing gain respectively in literature. Especially when the channel state information (CSI) is available, MIMO transceiver optimization can be conducted to greatly improve the communication quality \cite{JianYang,AnnaTIT,AnnaTSP1999,AnnaTSP2002}. In the last decades, MIMO technology has been mastered comprehensively and extensively with humongous amount of research results in the area.

One critical and initial question in MIMO designs that affects both the design approach and the resulting performance is how to choose the performance metric. Popular metrics include the capacity (or achievable rate), the mean-squared-error (MSE), the signal-to-noise-plus-interference ratio (SINR), the bit error rate (BER), and the outage probability. For single-input-single-output (SISO) systems, the selection of the performance metric is straightforward since they are usually equivalent to each other, thus the optimization of one leads to the optimization of all others. Unfortunately, for MIMO transceiver designs, the problem is considerably more complicated\cite{Palomar03}, due to the existence of multiple parallel sub-channels and multiple data streams. 
While these different performance metrics are still related with each other, they have critical differences. For example, to minimize the outage probability, more resources should be allocated to worse sub-channels to achieve balanced reliability across the channels, while for the sum-rate  or capacity maximization, more resources should be allocated to better sub-channels \cite{Tse05}.

It is thus highly desirable to find a
unified model for different performance metrics, which can lead to systematic MIMO transceiver optimization and help reveal fundamental connections among different performance metrics. As explained above, the challenge lies in the multiple sub-channels and data streams in MIMO communications, which makes the MIMO transceiver design a \textit{multi-objective optimization problem} \cite{XingTSP201501}. Motivated by the great success of scalarization in multi-objective optimization theory, the weighed MSE model is widely accepted as a powerful and effective way to unify different performance metrics. By leveraging the values of the weights \cite{YCEldar,Sampth01}, this model can achieve the optimization of other performance metrics such as the BER, or reach a desired balance among different performance metrics. The wide acceptance of MSE is based on the two facts. Firstly, sum-MSE is a popular metric that reflects how accurately signals can be recovered \cite{Joham05,Schizas07,Guan08}.
It plays an important role in communication, radar, sonar, and many other information systems \cite{Kay93}. Secondly, weighted MSE has a quadratic formula that is much easier to deal with than many other metrics \cite{XingIET}.
With the weighted MSE model, the MIMO\ transceiver design is formulated into finding the MIMO transceiver matrix that minimizes the weighted MSE under some resource constraints, usually the power resource. To the best of the authors' knowledge, there are two major and successful methodologies to solve such problem.

The first one uses the Lagrange multiplier method \cite{Sampth01}. In this method, the Karush-Kuhn-Tucker (KKT) conditions are firstly derived via complex matrix derivatives, from which the singular matrices of the MIMO transceiver matrix are obtained. Then with the help of the obtained results on the singular value decomposition (SVD) structure of the MIMO transceiver matrix, the optimization problem is largely simplified and the singular values  of the MIMO transceiver matrix (corresponding to the power allocation over sub-channels or data streams) are solved \cite{Serbetli04}.  This method has a wide range of applications with both perfect and imperfect CSI \cite{XingTSP2010}.

The second methodology for weighted MSE minimization is based on  majorization theory \cite{Palomar03}, an important branch of matrix inequality theory. It was shown in \cite{Palomar03} that for many setups of the weighted MSE minimization, when the objective function is shown to be Schur-concave or Schur-convex, the optimal structure of the MIMO transceivers can be derived. Similar to the first methodology, the structure  can greatly simplify the optimization problems via dimension reduction.  Majorization theory is less popular in MIMO designs compared to the Lagrange-multiplier method. However when it can be applied, it usually has simpler derivations. In addition, the large number of lemmas and theorems in majorization theory make the method highly potential to be further explored in MIMO\ designs.

Recently, a new extension for the weighed MSE model was proposed in \cite{XingComletter}, namely the matrix-filed weighted MSE minimization, where the weighting operation in the model is generalized from a vector operation (an inner product between the vector consisting of weighting factors and that consisting of the diagonal elements of the MSE matrix) to a general matrix operation. A distinct and important  contribution of this model is that it applies to the more complicated dual-hop amplify-and-forward (AF) MIMO relaying systems \cite{Tang07,Guan08,Medina07} and can represent their transceiver optimization problem as a matrix-field weighted MSE problem in point-to-point MIMO systems \cite{XingComletter}. This discovery helps uncover the answer to the question: why most solutions for point-to-point MIMO transceiver designs can be extended to the corresponding dual-hop AF MIMO relaying systems.

This paper is on the weighted MSE model and its optimization schemes for MIMO communications. The motivations to work on this well-investigated topic  are three-fold. Firstly, the topic is fundamental for MIMO communications and MIMO is becoming a key ingredient of current and future communication systems such as cognitive radio, cooperative communications, heterogeneous networks, wireless cloud networks, etc. Thus the topic is well worthy of an systematic and in-depth review. Second, while weighted MSE minimization is widely used in many papers, its appearance is somewhat sporadic. The fundamental ideas and optimization framework are buried in complicated and lengthy mathematical derivations. Moreover, in most existing work, the weaknesses and limitations of the existing methods are  overlooked  or glossed over, some of  which are critical for MIMO designs and need to be clarified. In this work, with rigorous mathematical structure, we elucidate the weaknesses and limitations of the methods and explain the problems they may cause. This is to help appropriate use of the model in future research and applications. Lastly, while weighted MSE model has been widely used, most work is limited to specific performance metrics via adjusting the balance among the MSEs of different data streams. With the new matrix-field weighted MSE model, we investigate its potential in more complicated scenarios such as nonlinear transceiver designs and the capacity maximization for general MIMO networks.

We would like to note that this work is inspired by the valuable and professional comments the authors received from anonymous reviewers on their journal submissions papers over the past 5 years. As some questions have been asked again and again, after carefully considering these questions and noticing the lack of proper and clear treatments on the issues, the authors believe that a rigorous and systematic review on weighted MSE model for MIMO communications is well deserved. More importantly, the limitations of existing methods should be emphasized. Some problems
that have been overlooked in existing work are fundamental and of critical importance even for future MIMO research.

The main contributions of our work are listed as follows.
\begin{itemize}
\item We provide a systematic and rigorous review on the two major methods to solve the weighted MSE minimization problem: the Lagrange multiplier method and the majorization theory based method.

\item For the Lagrange multiplier method, we reveal an ambiguity problem in the procedure of finding the right singular matrix of the MIMO precoding matrix from the KKT conditions. Existing procedure actually cannot determine the right singular matrix. In addition, there is a common belief that the KKT conditions lead to the water-filling solution for the power allocation. We point out that this claim is not for granted for the general case

\item For the majorization theory based method,  we show that the applicability of existing results on the SVD structure of the optimal solution does not solely rely on the Schur-convexity or Schur-concavity property of the objective function, but has strict limitations on the MSE matrix and power constraint formulations. This largely narrows down the applications of the majorization theory based method in MIMO transceiver designs, making it a supplement branch for the optimal solution derivation.

\item An extended matrix-field weighted MSE model is proposed in this paper. One important feature is that it can cover a wider range of MIMO\ designs, including capacity maximization and nonlinear transceiver designs with Tomlinson-Harashima precoding (THP) or decision feedback equalizer (DFE). Solutions to MIMO transceiver designs with the extended model are discussed.
\end{itemize}

\noindent {\textbf{Notation}}: Throughout this paper, the notation ${\bf{Z}}^{\rm{T}}$ and ${\bf{Z}}^{\rm{H}}$ denote the transpose and Hermitian transpose of matrix ${\bf{Z}}$, respectively and ${\rm{Tr}}({\bf{Z}})$ is the trace of matrix ${\bf{Z}}$. The symbol $\mathbb{E}\{\bullet\}$ represents the expectation operation. The matrix ${{\bf{Z}}^{\frac{1}{2}}}$ is the Hermitian square root of the positive semi-definite matrix ${\bf{Z}}$. The symbol ${a^ + }$ means $\max \{ 0,a\} $. The symbol $\lambda_i({\bf{Z}})$ denotes the $i^{\rm{th}}$ largest eigenvalue of matrix ${\bf{Z}}$. In addition, for two Hermitian matrices, the equation ${\bf{C}} \succeq
{\bf{D}}$ means that ${\bf{C}}-{\bf{D}}$ is a positive semi-definite
matrix. We use the expression ${\boldsymbol \Lambda} \searrow $ to represent a rectangular or square diagonal matrix with diagonal elements in decreasing order.

\section{MSE Minimization Problem Formulation for MIMO Systems}
\label{sec-model}
We consider a point-to-point MIMO system consisting of one source and one destination both equipped with multiple antennas.  The source transmits multiple data streams simultaneously. Denote the numbers of transmit antennas and receive antennas as $N_{\rm{Tx}}$ and $N_{\rm{Rx}}$, respectively. The corresponding transceiver model is
\begin{align}
{\bf{y}}={\bf{H}}{\bf{F}}{\bf{s}}+{\bf{n}},
\end{align}where ${\bf{y}}$ is the $N_{\rm{Rx}}\times 1$ received signal at the destination. The matrix ${\bf{H}}$ is the $N_{\rm{Rx}}\times N_{\rm{Tx}}$ channel matrix between the transmitter and the receiver. Moreover, ${\bf{s}}$ is the $N_{\rm{Dat}}\times 1$ transmitted signal vector, normalized as ${\mathbb E}\{{\bf{s}}{\bf{s}}^{\rm{H}}\}={\bf{I}}$, and ${\bf{F}}$ is the precoding matrix at the transmitter. Finally, the $N_{\rm{Rx}}\times 1$ vector ${\bf{n}}$ denotes the additive Gaussian noise vector at the receiver with zero mean and covariance matrix of ${\bf{R}}_n$.

At the receiver, by exploiting linear equalizer ${\bf{G}}$ to recover the desired signals, the MSE matrix of the data detection equals
\begin{align}
\label{MSE_Matrix}
{\boldsymbol \Phi}_{\rm{MSE}}({\bf{G}},{\bf{F}})= \mathbb{E}\{({\bf{G}}{\bf{y}}-{\bf{s}})
({\bf{G}}{\bf{y}}-{\bf{s}})^{\rm{H}}\}.
\end{align}
The diagonal elements of the MSE matrix ${\boldsymbol \Phi}$ correspond the SINRs of the data streams. Therefore, the MSE minimization problem can be formulated as
\begin{align}
\label{class_opt}
& \min_{{\bf{F}}} \ \ {\bf{d}}\left[{\boldsymbol \Phi}_{\rm{MSE}}({\bf{G}},{\bf{F}})\right]\nonumber \\
& \ \  {\rm{s.t.}} \ \ g_k({\bf{F}}) \le 0 \ \ k=1,\cdots, K,
\end{align}where ${\bf{d}}\left[{\bf{Z}}\right]$ denotes the vector consisting of the diagonal elements of ${\bf{Z}}$, that is, ${\bf{d}}\left[{\bf{Z}}\right]=[[{\bf{Z}}]_{1,1},\cdots,[{\bf{Z}}]_{N,N}]^{\rm{T}}$. The constraints $ g_k({\bf{F}}) \le 0$ correspond to the power constraints which can take different forms, e.g., the sum power constraint, the per-antenna power constraint, the shaping power constraint, etc. Note that the problem in  (\ref{class_opt}) is a multi-objective optimization problem whose objective functions are the MSEs of the $N$ signals \cite{Palomar03}.

It has been shown that the optimal equalizer is the linear minimum mean square error (LMMSE) equalizer:
\begin{align}
{\bf{G}}_{\rm{LMMSE}}=({\bf{H}}{\bf{F}})^{\rm{H}}({\bf{H}}{\bf{F}}
{\bf{F}}^{\rm{H}}{\bf{H}}^{\rm{H}}+{\bf{R}}_{n})^{-1}
\end{align}as it satisfies the following property \cite{XingTSP2015}
\begin{align}
{\boldsymbol \Phi}_{\rm{MSE}}({\bf{G}}_{\rm{LMMSE}},{\bf{F}}) \preceq {\boldsymbol \Phi}_{\rm{MSE}}({\bf{G}},{\bf{F}}).
\end{align}With the LMMSE equalizer, from (\ref{MSE_Matrix}), the MSE matrix reduces to:
\begin{align}
\label{Matrix_MSE}
{\boldsymbol \Phi}_{\rm{MSE}}({\bf{G}}_{\rm{LMMSE}},{\bf{F}})=({\bf{F}}^{\rm{H}}{\bf{H}}^{\rm{H}}{\bf{R}}_{n}^{-1}{\bf{H}}{\bf{F}}
+{\bf{I}})^{-1}.
\end{align}The optimization problem in (\ref{class_opt}) can be simplified as follows
\begin{align}
\label{OPT_Problem}
& \min_{{\bf{F}}} \ \ {\bf{d}}\left[({\bf{F}}^{\rm{H}}{\bf{H}}^{\rm{H}}{\bf{R}}_{n}^{-1}{\bf{H}}{\bf{F}}
+{\bf{I}})^{-1}\right]\nonumber \\
& \ \  {\rm{s.t.}} \ \ g_k({\bf{F}}) \le 0 \ \ k=1,\cdots, K.
\end{align} Then the remaining problem becomes how to solve (\ref{OPT_Problem}).

It is noteworthy that for single antenna systems, since only single data stream is transmitted, the problem becomes the traditional single-objective one which minimizes the MSE of the transmitted signal. Moreover, the  MSE minimization is equivalent to the SINR maximization \cite{Cover} as
\begin{align}
{\rm{MSE}}=\frac{1}{1+{\rm{SINR}}}.
\end{align}Unfortunately, in the case of MIMO systems with more than one data stream, the problem is totally different.
For the multi-objective optimization problem in (\ref{OPT_Problem}), scalarization is an effective method, which results in the popular weighted MSE minimization model, i.e., to minimize the weighted sum of the diagonal elements of ${\boldsymbol \Phi}_{\rm{MSE}}({\bf{G}}_{\rm{LMMSE}},{\bf{F}})$. By introducing a diagonal weight matrix ${\boldsymbol \Lambda}_{{\bf{w}}}$ the weighted MSE minimization problem can be succinctly rewritten as
\begin{align}
\label{OPT_Original_1}
& \min_{{\bf{F}}} \ \ \ \ {\rm{Tr}}[{\boldsymbol \Lambda}_{{\bf{w}}}({\bf{F}}^{\rm{H}}{\bf{H}}^{\rm{H}}{\bf{R}}_{n}^{-1}{\bf{H}}{\bf{F}}
+{\bf{I}})^{-1}] \nonumber \\ & \  {\rm{s.t.}} \ \ \ \ \  g_k({\bf{F}}) \le 0 \ \ k=1,\cdots, K.
\end{align}

Without loss of generality, denote $w_i$ to be the $i^{\rm{th}}$ largest diagonal entry of  ${\boldsymbol \Lambda}_{{\bf{w}}}$.
In existing work the logic to solve the above matrix-variable optimization problem consists of two stages. In the first stage, results on the structure (e.g., the SVD structure) of the optimal precoding matrix are derived, based on which the unknown variable is simplified from a general matrix to a diagonal real matrix (e.g., the singular value matrix of the precoding matrix). In the second stage, the optimal solution of the diagonal elements is derived and usually claimed to be the water-filling solution. For the first stage, two common methods in the existing literature are the  Lagrange multiplier method and the majorization theory based method. In general, if both are applicable, the later one has tidier derivations than the former; while the later one has more strict limitations in applications. For both methods, some critical facts are often ignored in exiting work. In this work, we clarify these facts and point out limitations of the methods.

%

\section{Lagrange Multiplier Method}
In this section, we review the Lagrange multiplier method for the weighted MSE minimization problem in MIMO transceiver design. Then the existence of the permutation ambiguity effect and the turning-off effect of this method are pointed out. These issues bring obstacles in finding of the optimal solution via this method.

\subsection{Fundamentals of Lagrange Multiplier Method}
Lagrange multiplier method and its KKT conditions are widely used for solving optimization problems in wireless communications \cite{Sampth01}. When certain regularity conditions are satisfied, KKT conditions are necessary conditions for the optimal solution and they can provide important information such as the optimal structure of the matrix variable. Then the information is exploited further to derive the optimal solution. Generally speaking, Lagrange multiplier method consists of the following four steps.
\begin{itemize}
\item \textbf{Step 1)} Based on the original optimization problem, its Lagrangian is formulated by introducing the Lagrange multipliers.

\item \textbf{Step 2)} By calculating matrix derivatives (especially complex matrix derivatives), the KKT conditions are derived.

\item \textbf{Step 3)} Based on the KKT conditions, the optimal structure of the matrix variable is derived, typically the left and right singular matrices of the matrix variable.

\item \textbf{Step 4)} Based on the derived structure, the optimization problem can be greatly simplified into a simpler one with a series of scalar variables, typically the singular values of the matrix variable.
\end{itemize}

A key point of this method lines in Step 3 on how to take advantage of the KKT conditions to derive the optimal structure of the matrix variables. The derivations are usually based on the following facts.

\noindent \textbf{Theorem 1 \cite{Horn85}:} For two Hermitian matrices ${\bm{A}}$ and ${\bm{B}}$ when ${\bm{A}}{\bm{B}}={\bm{B}}{\bm{A}}$ it can be concluded that there exists a unitary matrix that can diagonalize ${\bm{A}}$ and ${\bm{B}}$ simultaneously.

\noindent \textbf{Theorem 2 \cite{Horn85}:} For a rectangular matrix ${\bm{A}}$, the EVD unitary matrices of  ${\bm{A}}{\bm{A}}^{\rm{H}}$ and ${\bm{A}}^{\rm{H}}{\bm{A}}$ are the left and right SVD unitary matrices of ${\bm{A}}$, respectively.

\subsection{A Specific Application}

In this section, a specific application is used to exhibit the Lagrange multiplier based method for the precoding matrix optimization. We focus on a simple special case of (\ref{OPT_Original_1}), where the sum power constraint is considered\footnote{We would like to clarify that the detailed procedure of solution we present in the following is different from the existing work in \cite{Sampth01}, but the ideas are fundamentally the same. The difference lies in the fact that in our method the equalizer ${\bf{G}}$ has been analytically solved as by a function of ${\bf{F}}$. Our method is clearer as there are less variables.
}. The optimization problem has the following form
\begin{align}
\label{OPT_Original}
& \min_{{\bf{F}}} \ \ \ \ {\rm{Tr}}[{\boldsymbol \Lambda}_{{\bf{w}}}({\bf{F}}^{\rm{H}}{\bf{H}}^{\rm{H}}{\bf{R}}_{n}^{-1}{\bf{H}}{\bf{F}}
+{\bf{I}})^{-1}] \nonumber \\ & \  {\rm{s.t.}} \ \ \ \ \ {\rm{Tr}}({\bf{F}}{\bf{F}}^{\rm{H}}) \le P.
\end{align}

As Step 1, the Lagrangian of (\ref{OPT_Original}) is written as
\begin{align}
{\mathcal {L}}({\bf{F}},\mu)={\rm{Tr}}[{\boldsymbol {\Lambda}}_{{\bf{w}}}({\bf{F}}^{\rm{H}}{\bf{H}}^{\rm{H}}{\bf{R}}_{n}^{-1}{\bf{H}}{\bf{F}}
+{\bf{I}})^{-1}]+\mu[{\rm{Tr}}({\bf{F}}{\bf{F}}^{\rm{H}})-P],
\end{align}where the real scalar $\mu$ is the lagrange multiplier. For Step 2, by taking the derivatives of ${\mathcal {L}}({\bf{F}},\mu)$ with respect to $\bf{F}$ and $\mu$ and making them 0, the  KKT conditions can be derived to be \cite{Boyd04}
\begin{align}
\label{KKT_Conditions}
& {\bf{H}}^{\rm{H}}{\bf{R}}_{\bf{n}}^{-1}{\bf{H}}{\bf{F}}({\bf{F}}^{\rm{H}}{\bf{H}}^{\rm{H}}{\bf{R}}_{\bf{n}}^{-1}{\bf{H}}{\bf{F}}
+{\bf{I}})^{-1}{\boldsymbol {\Lambda}}_{{\bf{w}}}\nonumber\\
& \ \ \ \ \ \ \ \ \ \ \ \ \ \times({\bf{F}}^{\rm{H}}{\bf{H}}^{\rm{H}}{\bf{R}}_{\bf{n}}^{-1}{\bf{H}}{\bf{F}}
+{\bf{I}})^{-1} =\mu {\bf{F}}. \nonumber  \\
& \mu({\rm{Tr}}({\bf{F}}{\bf{F}}^{\rm{H}})-P)=0. \nonumber  \\
 & \mu \ge 0, \ \  {\rm{Tr}}({\bf{F}}{\bf{F}}^{\rm{H}})\le P.
\end{align}

The next step, Step 3, is to derive the optimal structure of $\bf{F}$. Based on the first KKT condition, right multiplying ${\bf{F}}^{\rm{H}}$ we have the following equation
\begin{align}
\label{KKT_1_1}
&  {\bf{H}}^{\rm{H}}{\bf{R}}_{\bf{n}}^{-1}{\bf{H}}{\bf{F}}
({\bf{F}}^{\rm{H}}{\bf{H}}^{\rm{H}}{\bf{R}}_{\bf{n}}^{-1}{\bf{H}}{\bf{F}}
+{\bf{I}})^{-1}\nonumber \\
& \ \ \ \ \ \ \times{\boldsymbol {\Lambda}}_{{\bf{w}}}({\bf{F}}^{\rm{H}}{\bf{H}}^{\rm{H}}{\bf{R}}_{\bf{n}}^{-1}{\bf{H}}{\bf{F}}
+{\bf{I}})^{-1}{\bf{F}}^{\rm{H}} =\mu{\bf{F}}{\bf{F}}^{\rm{H}}.
\end{align}
It is obvious that on the right-hand side of this equation, $\mu{\bf{F}}{\bf{F}}^{\rm{H}}$ is a Hermitian matrix. In addition, on the left-hand side ${\bf{H}}^{\rm{H}}{\bf{R}}_{\bf{n}}^{-1}{\bf{H}}$ and ${\bf{F}}
({\bf{F}}^{\rm{H}}{\bf{H}}^{\rm{H}}{\bf{R}}_{\bf{n}}^{-1}{\bf{H}}{\bf{F}}
+{\bf{I}})^{-1}{\boldsymbol {\Lambda}}_{{\bf{w}}}({\bf{F}}^{\rm{H}}{\bf{H}}^{\rm{H}}{\bf{R}}_{\bf{n}}^{-1}{\bf{H}}{\bf{F}}
+{\bf{I}})^{-1}{\bf{F}}^{\rm{H}}$ are both Hermitian matrices. It can be shown easily that if the product of two Hermitian matrices is Hermitian, then the two matrices commute. Based on \textbf{Theorem 1}, ${\bf{H}}^{\rm{H}}{\bf{R}}_{\bf{n}}^{-1}{\bf{H}}$ and ${\bf{F}}
({\bf{F}}^{\rm{H}}{\bf{H}}^{\rm{H}}{\bf{R}}_{\bf{n}}^{-1}{\bf{H}}{\bf{F}}
+{\bf{I}})^{-1}{\boldsymbol {\Lambda}}_{{\bf{w}}}({\bf{F}}^{\rm{H}}{\bf{H}}^{\rm{H}}{\bf{R}}_{\bf{n}}^{-1}{\bf{H}}{\bf{F}}
+{\bf{I}})^{-1}{\bf{F}}^{\rm{H}}$ can be simultaneously diagonalized by the same unitary matrix, meaning that the two have the same eigenvalue decomposition (EVD) unitary matrix.
Further from the equality in (\ref{KKT_1_1}), it can be shown that ${\bf{H}}^{\rm{H}}{\bf{R}}_{\bf{n}}^{-1}{\bf{H}}$ and $\mu{\bf{F}}{\bf{F}}^{\rm{H}}$ have the same EVD unitary matrix. Then based on \textbf{Theorem 2}, it can be concluded that \textit{the left SVD unitary matrix of ${\bf{F}}$ is the EVD unitary matrix of ${\bf{H}}^{\rm{H}}{\bf{R}}_{\bf{n}}^{-1}{\bf{H}}$}.

On the other hand, based on the first KKT condition, left-multiplying both sides of the equation with ${\bf{F}}^{\rm{H}}$, we have the following equality
\begin{align}
& {\bf{F}}^{\rm{H}} {\bf{H}}^{\rm{H}}{\bf{R}}_{\bf{n}}^{-1}{\bf{H}}{\bf{F}}({\bf{F}}^{\rm{H}}{\bf{H}}^{\rm{H}}{\bf{R}}_{\bf{n}}^{-1}{\bf{H}}{\bf{F}}
+{\bf{I}})^{-1}\nonumber \\
& \ \ \ \ \ \ \times{\boldsymbol {\Lambda}}_{{\bf{w}}}({\bf{F}}^{\rm{H}}{\bf{H}}^{\rm{H}}{\bf{R}}_{\bf{n}}^{-1}{\bf{H}}{\bf{F}}
+{\bf{I}})^{-1} =\mu {\bf{F}}^{\rm{H}}{\bf{F}}.
\end{align}
Thus,
\begin{align}
&({\bf{F}}^{\rm{H}}{\bf{H}}^{\rm{H}}{\bf{R}}_{\bf{n}}^{-1}{\bf{H}}{\bf{F}}
+{\bf{I}}) {\bf{F}}^{\rm{H}} {\bf{H}}^{\rm{H}}{\bf{R}}_{\bf{n}}^{-1}{\bf{H}}{\bf{F}}\nonumber \\
&  \ \ \ \ \ \ \ \ \ \ \ \ \ \ \ \ \times({\bf{F}}^{\rm{H}}{\bf{H}}^{\rm{H}}{\bf{R}}_{\bf{n}}^{-1}{\bf{H}}{\bf{F}}
+{\bf{I}})^{-1}{\boldsymbol {\Lambda}}_{{\bf{w}}}\nonumber \\
=& \mu ({\bf{F}}^{\rm{H}}{\bf{H}}^{\rm{H}}{\bf{R}}_{\bf{n}}^{-1}{\bf{H}}{\bf{F}}
+{\bf{I}}){\bf{F}}^{\rm{H}}{\bf{F}} ({\bf{F}}^{\rm{H}}{\bf{H}}^{\rm{H}}{\bf{R}}_{\bf{n}}^{-1}{\bf{H}}{\bf{F}}
+{\bf{I}}).
\label{eqn-14}
\end{align}
By following similar arguments and noticing the special structure of the matrices in (\ref{eqn-14}), it can be proved that ${\bf{F}}^{\rm{H}}{\bf{H}}^{\rm{H}}{\bf{R}}_{\bf{n}}^{-1}{\bf{H}}{\bf{F}}$ and ${\boldsymbol {\Lambda}}_{{\bf{w}}}$ can be simultaneously diagonalized. Together with the fact that the left SVD unitary matrix of ${\bf{F}}$ is the EVD unitary matrix of ${\bf{H}}^{\rm{H}}{\bf{R}}_{\bf{n}}^{-1}{\bf{H}}$, it can be concluded that \textit{the right SVD unitary matrix of ${\bf{F}}$ is the EVD unitary matrix of ${\boldsymbol {\Lambda}}_{{\bf{w}}}$}.

From the derived results, the following conclusion on the SVD strcuture of ${\bf F}$ is obtained.
\begin{conclusion}
Let the SVD of ${\bf{R}}_{\bf{n}}^{-1/2}{\bf{H}}$ and the EVD of ${{\boldsymbol \Lambda}_{\bf{w}}}$ be as follows
\begin{eqnarray}
&& \label{channel_decomp}
{\bf{R}}_{\bf{n}}^{-1/2}{\bf{H}}=
{\bf{U}}_{{\boldsymbol{\mathcal{H}}}}{\boldsymbol \Lambda}_{{\boldsymbol{\mathcal{H}}}}{\bf{V}}_{{\boldsymbol{\mathcal{H}}}}
^{\rm{H}} \ \ {\rm{with}} \ \ {\boldsymbol \Lambda}_{{\boldsymbol{\mathcal{H}}}} \searrow,\\
&&\label{diagonal_decom}
{{\boldsymbol \Lambda}_{\bf{w}}}={\bf{U}}_{\rm{Per}}{{\boldsymbol{\tilde \Lambda}}_{\bf{w}}}{\bf{U}}_{\rm{Per}}^{\rm{H}}.
\end{eqnarray}
The matrix variable ${\bf{F}}$ satisfying the KKT conditions owns the following structure
\begin{align}
{\bf{F}}={\bf{V}}_{{\boldsymbol{\mathcal{H}}}}{\boldsymbol\Lambda}_{\bf{F}}
{\bf{U}}_{\rm{Per}}^{\rm{H}},
\label{F-decomp}
\end{align}
where ${\boldsymbol\Lambda}_{\bf{F}}$ is a rectangular diagonal matrix. 
\end{conclusion}

With this SVD structure, the last step is conducted by adopting (\ref{F-decomp}) in the first KKT condition in (\ref{KKT_Conditions}), which leads to the following equality:
\begin{align}
\label{KKT_1_2}
{\boldsymbol{\Lambda}}_{\bf{F}}^{\rm{T}}{\boldsymbol \Lambda}_{{\boldsymbol{\mathcal{H}}}}^{\rm{T}}{\boldsymbol \Lambda}_{{\boldsymbol{\mathcal{H}}}}{\boldsymbol{\Lambda}}_{\bf{F}}
({\boldsymbol{\Lambda}}_{\bf{F}}^{\rm{T}}{\boldsymbol \Lambda}_{{\boldsymbol{\mathcal{H}}}}^{\rm{T}}{\boldsymbol \Lambda}_{{\boldsymbol{\mathcal{H}}}}{\boldsymbol{\Lambda}}_{\bf{F}}+{\bf{I}})^{-1}
{\bf{U}}_{\bf{per}}^{\rm{H}}{\boldsymbol {\Lambda}}_{{\bf{w}}}{\bf{U}}_{\bf{Per}}({\boldsymbol{\Lambda}}_{\bf{F}}^{\rm{T}}
{\boldsymbol \Lambda}_{{\boldsymbol{\mathcal{H}}}}^{\rm{T}}{\boldsymbol \Lambda}_{{\boldsymbol{\mathcal{H}}}}{\boldsymbol{\Lambda}}_{\bf{F}}+{\bf{I}})^{-1}
=\mu{\boldsymbol{\Lambda}}_{\bf{F}}^{\rm{T}}{\boldsymbol{\Lambda}}_{\bf{F}}.
\end{align}By noticing that ${\bf{U}}_{\bf{per}}^{\rm{H}}{\boldsymbol {\Lambda}}_{{\bf{w}}}{\bf{U}}_{\bf{Per}}={\boldsymbol {\tilde \Lambda}}_{{\bf{w}}}$ is still a diagonal matrix, it was claimed in existing literature that the diagonal matrix ${\boldsymbol{\Lambda}}_{\bf{F}}$ can be solved from  (\ref{KKT_1_2}) and the result is the water-filling solution.

The method has plausible logic and sleek techniques. However, it has a couple of faults in finding the optimal precoding matrix, which we will point out in the next subsection.

\subsection{Comments on Lagrange Multiplier Method}

\subsubsection{\textbf{Permutation Ambiguous Effect}}

Although Conclusion 1 provides the optimal SVD structure of the precoding matrix, the matrix ${\bf{U}}_{\rm{Per}}$ cannot be uniquely determined from the KKT conditions. From (\ref{diagonal_decom}), as ${{\boldsymbol \Lambda}_{\bf{w}}}$ is diagonal, the unitary matrix ${\bf{U}}_{\rm{Per}}$  in general can be any permutation matrix that has one entry of 1 in each row and each column and 0's elsewhere\footnote{If ${{\boldsymbol \Lambda}_{\bf{w}}}$ has repeated diagonal entries,  ${\bf{U}}_{\rm{Per}}$ can take more general form. For the extreme case that ${{\boldsymbol \Lambda}_{\bf{w}}}$ is a multiple of ${\bf I}$, ${\bf{U}}_{\rm{Per}}$ can be any unitary matrix.}.
This comes from the fact that for EVD, the eigenvalues can be arranged in an arbitrary order. It can be shown straightforwardly that any unitary eigenmatrix of ${{\boldsymbol \Lambda}_{\bf{w}}}$ satisfies the KKT conditions. Different choices of  ${\bf{U}}_{\rm{Per}}$ will lead to different precoding matrix designs. As a result the optimal solution cannot be determined from the KKT conditions only, the claim that the optimal solution is found with Conclusion 1 is false. We call this the \textbf{permutation ambiguous effect} of the Lagrange multiplier method.

One way to determine the ${\bf{U}}_{\rm{Per}}$ matrix in Conclusion 1 for the optimal solution is via exhaustive search, where all possible ${\bf{U}}_{\rm{Per}}$ satisfying the KKT  conditions are checked by using their corresponding ${\bf F}$ in the objective function to find the optimal solution. But the computation complexity may be very high.
It is noteworthy that for the sum-MSE minimization, where ${{\boldsymbol \Lambda}_{\bf{w}}}={\bf{I}}$, the permutation ambiguous effect does not disappear but becomes more serious as ${\bf{U}}_{\rm{Per}}$ can be an arbitrary unitary matrix.

\noindent {\textbf{Comment 1:}} For the optimization problem in (\ref{OPT_Original}), it is impossible to determine ${\bf{U}}_{\rm{Per}}$ in Conclusion 1 solely based on the KKT conditions due to the permutation ambiguous effect.

\subsubsection{\textbf{Turning-Off Effect}}

Unfortunately, the permutation ambiguous effect is not the only fault in the method that results in more than one solution satisfying the KKT conditions. Another problem exists in finding the singular values of ${\bf F}$ in Step 4. In what follows, we show that even for a given ${\bf{U  }}_{\rm{Per}}$, the solution of (\ref{KKT_1_2}) is not guaranteed  to be the water-filling solution and its solution may not even be unique.

Let $[{\boldsymbol \Lambda}_{{\boldsymbol{\mathcal{H}}}}]_{i,i}=h_i$ and $[{\boldsymbol \Lambda}_{{\mathbf{F}}}]_{i,i}=f_i$. Since the diagonal entries of ${\boldsymbol \Lambda}_{{\boldsymbol{\mathcal{H}}}}$ are in decreasing order, $h_i$ is also the $i^{\rm{th}}$ largest diagonal entry of ${\boldsymbol \Lambda}_{{\boldsymbol{\mathcal{H}}}}$. With a given ${\bf{U}}_{\rm{Per}}$, the KKT conditions in (\ref{KKT_Conditions}) become
\begin{align}
\label{KKT_Conditions_Scalar}
&모\left\{\frac{f_ih_i^2w_i}{(1+f_i^2h_i^2)^2}=\mu f_i \right\}_{i=1}^N \nonumber \\
&  \mu(\sum_i f_i^2-P)=0, \nonumber \\
& \mu \ge 0, \ \ \sum_i f_i^2 \le P.
\end{align}
It cannot be concluded from the first equation of (\ref{KKT_Conditions_Scalar}) that \begin{align}
\frac{h_i^2w_1}{(1+f_i^2h_i^2)^2}=\mu \label{power-allo}
\end{align}
since $f_i$ may be zero. As a matter of fact, one can set any subset of $\{f_1,f_2,\cdots,f_N\}$ to takes the zero value, and allocate the power among the rest of the non-zero $f_i$'s according to (\ref{power-allo}) to obtain a solution of the KKT conditions. By having different subsets of zero $f_i$'s, different solutions are found. The water-filling solution can be obtained by having the right set of $f_i$'s to have zero value. But this cannot be obtained from the KKT conditions only. We name this the \textbf{turning-off effect} of the method.


To further clarify the difference to water-filling solution. We show the optimization problem, the KKT conditions, and the solution corresponding to water-filling. Consider the following convex optimization problem
\begin{align}
& \min_{p_i} \ \  \sum_{i=1}^N\frac{w_i}{1+p_ih_i^2} \nonumber\\
& \ {\rm{s.t.}} \ \ \sum_i p_i\le P, \ \ p_i \ge 0.
\end{align}The KKT conditions of the problem are as follows
\begin{align}
\label{KKT_Conditions_Convex}
& \left\{-\frac{w_ih_i^2}{(1+p_ih_i^2)^2}+\mu-\phi_i=0\right\}_{i=1}^N  \nonumber \\
& \mu \ge 0, \ \ \phi_i \ge 0, \ \ p_i \ge 0, \ \  \sum_i p_i\le P,
\end{align}
where $\mu$ is the Lagrange multiplier corresponding to the sum power constraint and $\phi_i$ is the Lagrange multiplier corresponding to the condition $p_i \ge 0$.
From the KKT conditions, the optimal solution is the famous water-filling solution in the following form
\begin{align}\label{WF-solution}
p_i=\left(\sqrt{\frac{w_i}{\mu h_i^2}}-\frac{1}{h_i^2}\right)^{+}.
\end{align}

Although for the water-filling solution, some eigenchannels are allocated zero powers (these eigenchannels are thus closed), the reason is the poor conditions of the channels and it is fundamentally different from the turning-off effect in (\ref{KKT_Conditions_Scalar}). Here we would like to highlight that the existence of the operation $+$ in (\ref{WF-solution}) is not because the power should be nonnegative so we enforce the negative values to be zeros. Although this logic coincides with our intuition, it cannot be used as a rigorous theoretical basis. As proved in \cite{Boyd04}, some eigenchannels are closed since their channel amplitudes are smaller than a threshold. The eigenchannels allocated with zeros are actually involved in the water-filling computation. The operation $+$ is simply to have a compact representation of the solution. Also, water-filling tends to occupy as many subchannels as possible. On the other hand, the turning-off effect does not have this property, nor does it consider the channel quality. From (\ref{KKT_Conditions_Scalar}), we see that one can freely choose the eigenchannels to be turned-off, even the ones with top channel conditions, and the resulting solution still satisfies the KKT conditions.


\noindent {\textbf{Comment 2:}} Due to the turning-off effect, it cannot be concluded from the KKT conditions only that the solution of the diagonal elements of ${\boldsymbol \Lambda}_{\bf{F}}$ in \textbf{Conclusion 1} is a water-filling solution.

Both the permutation ambiguous effect and the turning-off effect occur because the considered  optimization problem in (\ref{OPT_Original}) is not convex with respect to the matrix variable ${\bf{F}}$. Thus the corresponding KKT conditions are only necessary but not sufficient for the optimal solution. In previous work, some researchers used Lagrange multiplier method to derive the optimal solution based on the KKT conditions only, and indicated that this derivation is not affected by whether the optimization problem is convex or not. Unfortunately, based on our discussions, this idea is faulty. To avoid the drawbacks, the convexity of optimization problem needs to be carefully exploited. Unfortunately, only in some special cases the convexity of the problem holds or can be proved.

\subsubsection{\textbf{Relaxation Issues with Variable Transformation}}
In addition to the permutation-ambiguity effect and the turning-off effect, we would like to comment on the use of transformation in solving the MIMO transceiver optimization problem. Transformations need to be adopted with caution as improper ones can result in new issues such as rank constraint and so on.

For the sum-MSE minimization,  in some existing literature, the matrix ${\bf{F}}{\bf{F}}^{\rm{H}}$, instead of ${\bf{F}}$, is used as the variable of the optimization. The problem is formulated as
\begin{align}
\label{matrix_variable}
& \min_{{\bf{Q}}} \ \ {\rm{Tr}}\left[({\bf{R}}_{n}^{-1/2}{\bf{H}}{\bf{Q}}{\bf{H}}^{\rm{H}}{\bf{R}}_{n}^{-1/2}
+{\bf{I}})^{-1}\right]\nonumber \\
& \ \  {\rm{s.t.}} \ \ {\rm{Tr}}({\bf{Q}})\le P \nonumber \\
& \ \ \ \ \ \ \ \ {\bf{Q}} \succeq {\bf{0}}
\end{align}It can be shown that this optimization problem is convex with respect to ${\bf{Q}}$ and the optimal ${\bf{Q}}$ can be obtained using Lagrange multiplier method by working on the KKT conditions.  The eigenvalues of the optimal ${\bf{Q}}$ are a water-filling solution. However, the transformation of the matrix variable from ${\bf F}$ to  ${\bf Q}={\bf{F}}{\bf{F}}^{\rm{H}}$ to obtain the new optimization problem in (\ref{matrix_variable}) induces the following issues.

\begin{itemize}
\item The constraint on the rank of ${\bf{Q}}$ is relaxed. In (\ref{matrix_variable}), there is no constraint on the rank of ${\bf Q}$. In the original problem (\ref{OPT_Original}) (where ${\boldsymbol \Lambda}_{\bf{w}}={\bf I}$ for the sum-MSE minimization), the rank of ${\bf{F}}{\bf{F}}^{\rm{H}}$ is limited by the number of data streams, which can be smaller than the smaller number of the transmit and receive antennas. Generally speaking, rank constraint is nonconvex and the common scheme to deal with it is rank relaxation. However, in general, the solution of the transformed problem does not satisfy the rank constraint. Some further processing such as projection is needed to find a sub-optimal solution.

\item  This transformation does not work for weighted MSE minimization. It is worth highlighting that the following optimization problem is different from the original one in (\ref{OPT_Original})
\begin{align}
\label{matrix_variable_waighted}
& \min_{{\bf{Q}}} \ \ {\rm{Tr}}\left[{\boldsymbol \Lambda}_{\bf{w}}({\bf{R}}_{n}^{-1/2}{\bf{H}}{\bf{Q}}{\bf{H}}^{\rm{H}}{\bf{R}}_{n}^{-1/2}
+{\bf{I}})^{-1}\right]\nonumber \\
& \ \  {\rm{s.t.}} \ \ {\rm{Tr}}({\bf{Q}})\le P \nonumber \\
& \ \ \ \ \ \ \ \ {\bf{Q}} \succeq {\bf{0}}.
\end{align}The optimal solution of (\ref{matrix_variable_waighted}) does not satisfy the optimal structure given in Conclusion 1.
\end{itemize}

\subsection{Summary}

Lagrange multiplier method is easy to implement. Based on the KKT conditions, important structure of the optimal solution can be straightforwardly derived, which largely simplifies the derivations for the optimal solution. Its great success is due to the neat matrix derivatives and powerful matrix theory. Lagrange multiplier method has a wide range of applications in traditional MIMO transceiver designs, robust designs with random channel errors \cite{Ding09}, and even robust AF MIMO relaying systems \cite{XingTSP2010}.

Unfortunately, in most cases the KKT conditions are only necessary for the optimal solution. Based on the discussions above, we can see that the permutation-ambiguous effect and the turning-off effect both result in multiple solutions  satisfying the KKT conditions. In specific, the unitary matrix ${\bf{U}}_{\rm{per}}$  in Conclusion 1 cannot be determined purely relying on the KKT conditions. Furthermore, due to the turning-off effect, it cannot be claimed from the KKT conditions  that the optimal power allocation is water-filling. Actually, the KKT conditions can be satisfied by multiple power allocation solutions. Therefore, the optimal solution of the optimization problem cannot be found from the KKT conditions only.

\section{Majorization Theory Based Method}
In this section, the majorization theory based method for the weighted MSE minimization problem in MIMO transceiver design is reviewed. Then we point out that in the derived optimal structure, the permutation  matrix cannot be ignored.  The limitation of the method in various formulations of the weighted MSE minimization problem is also clarified.
\subsection{Fundamentals of Majorization Theory}

\noindent \textbf{Definition 1 \cite{Marshall79}:} For ${\bf{x}},{\bf{y}}$ $\in \mathcal{R}^{N}$, vector ${\bf{x}}$ is majorized by vector ${\bf{y}}$, denoted as ${\bf{x}} \prec {\bf{y}}$, if the following equations hold:
\begin{align}
& \sum_{i=1}^kx_{[i]}\le \sum_{i=1}^ky_{[i]},  \ \ \ \ \ \ \mbox{ for }  k=1,\cdots,N-1 \nonumber \\
& \sum_{i=1}^Nx_{[i]}=\sum_{i=1}^Ny_{[i]}.
\end{align}

\noindent \textbf{Definition 2 \cite{Marshall79}:} A real-valued function $\phi$ defined on a feasible set is Schur-convex if for any ${\bf x},{\bf y}$ in the feasible set,
\begin{align}
{\bf{x}} \prec {\bf{y}} \rightarrow \phi({\bf{x}})\le \phi({\bf{y}}).
\end{align} On the other hand, $\phi$ is Schur-concave if for any ${\bf x},{\bf y}$ in the feasible set,
\begin{align}
{\bf{x}} \prec {\bf{y}} \rightarrow \phi({\bf{x}})\ge \phi({\bf{y}}).
\end{align}

Majorization theory can be a powerful mathematical tool for MIMO transceiver design problems. When applicable, it can help derive the optimal structure of the matrix variable and avoid the ambiguity effect of the Lagrange multiplier method.

\subsection{Majorization Theory Based Method}
D. P. Palomar et al. have proposed a framework for MIMO transceiver optimization using majorization theory \cite{Palomar03}. Majorization theory based method is less popular in the literature of MIMO design and may  look mysterious for some wireless engineers. Simplify speaking, the role of majorization theory is to derive the optimal structure of matrix variable, similar to Step 3 of the Lagrange multiplier method but without the permutation  ambiguity as it can specify the relative orders of the eigenvalues of the matrices in the problem.
In this subsection, we use two examples to show the common procedure of majorization theory based method.

\noindent \textbf{Case 1: Training Sequence Optimization}

For the MIMO system described in Section \ref{sec-model}, the training sequence designs with correlated signal and colored noise can be formulated as \cite{Liu2007}
\begin{align}
\label{training-prob}
& \min_{{\bf{X}}_{\rm{T}}} \ \ {\rm{Tr}}\left[({\bf{X}}_{\rm{T}}{\bf{R}}_{n}^{-1}{\bf{X}}_{\rm{T}}^{\rm{H}}
+{\bf{R}}_{\bf{H}}^{-1})^{-1}\right]\nonumber \\
& \ \  {\rm{s.t.}} \ \ {\rm{Tr}}({\bf{X}}_{\rm{T}}^{\rm{H}}{\bf{X}}_{\rm{T}})\le P,
\end{align} where ${\bf{R}}_{n}$ and ${\bf{R}}_{\bf{H}}$ are positive definite matrices representing the covariance of the noise vector and channel matrix, respectively.
In the existing work \cite{Liu2007}, this optimization problem is solved by the Lagrange multiplier method with complicated derivations. Here, a much simpler procedure to obtain the solution is provided using majorization theory.

Notice that the function $1/x$ is Schur-convex for $x>0$ and  ${\boldsymbol \lambda}( {\bm{A}} +{\bm{B}})$ majorizes ${\boldsymbol \lambda}( {\bm{A}})\uparrow +{\boldsymbol \lambda}({\bm{B}}) \downarrow$ \cite{Jorswieck07}, where ${\boldsymbol \lambda}( {\bm{Z}})\uparrow $ and ${\boldsymbol \lambda}( {\bm{Z}})\downarrow $ denote the vectors consisting of the eigenvalues of ${\bm{Z}}$ in increasing order and decreasing order, separately. We can directly conclude that the optimal solution of (\ref{training-prob}) has the following structure
\begin{align}
{\bf{X}}_{\rm{T}}={\bf{U}}_{{\bf{R}}_n}{\boldsymbol\Lambda}_{\bf{X}}{\bf{\tilde U}}_{{\bf{R}}_{\rm{H}}}^{\rm{H}},
\label{training-stru}
\end{align} where ${\boldsymbol\Lambda}_{\bf{X}}$ is a diagonal matrix and the unitary matrices ${\bf{U}}_{{\bf{R}}_n}$ and ${\bf{U}}_{{\bf{R}}_H}$ are defined based on the following EVDs
\begin{align}
{\bf{R}}_n^{-1}&={\bf{U}}_{{\bf{R}}_n}{ \boldsymbol{ \Lambda}}_{{\bf{R}}_n}{\bf{U}}_{{\bf{R}}_n}^{\rm{H}} \ \ {\rm{with}} \ \ { \boldsymbol{ \Lambda}}_{{\bf{R}}_n} \searrow \\
{\bf{R}}_{\rm{H}}^{-1}&={\bf{\tilde U}}_{{\bf{R}}_{\rm H}}{ \boldsymbol{\tilde \Lambda}}_{{\bf{R}}_{\rm H}}{\bf{\tilde U}}_{{\bf{R}}_{\rm H}}^{\rm{H}} \ \ {\rm{with}} \ \ { \boldsymbol{\tilde \Lambda}}_{{\bf{R}}_{\rm H}} \nearrow.
\end{align}The opposite orderings of the eigenvalues of the two EVDs are very important for the result in (\ref{training-stru}). Given the optimal structure in (\ref{training-stru}), the diagonal matrix ${\boldsymbol\Lambda}_{\bf{X}}$ can be further computed, for example by using the the KKT conditions as explained in the previous section.

\noindent \textbf{Case 2: MIMO Transceiver Design with Weighted MSE Minimization under Sum-Power Constraint }

The formulation of the weighted MSE minimization under sum-power constraint is given in (\ref{OPT_Original}).
In the framework proposed in \cite{Palomar03}, it was shown that when the diagonal elements of ${\boldsymbol \Lambda}_{{\bf{w}}}$ are in decreasing order and the diagonal elements of the MSE matrix $({\bf{F}}^{\rm{H}}{\bf{H}}^{\rm{H}}{\bf{R}}_{n}^{-1}{\bf{H}}{\bf{F}}
+{\bf{I}})^{-1}$ are in increasing order, the objective function of the weighted MSE minimization problem can be understood as a Schur-concave function. More specifically, the function $f(\cdot)$ defined as
\begin{align}
f({\bf{d}}\{({\bf{F}}^{\rm{H}}{\bf{H}}^{\rm{H}}{\bf{R}}_{n}^{-1}{\bf{H}}{\bf{F}}
+{\bf{I}})^{-1}\})=
{\rm{Tr}}\left[{\boldsymbol \Lambda}_{{\bf{w}}}({\bf{F}}^{\rm{H}}{\bf{H}}^{\rm{H}}{\bf{R}}_{n}^{-1}{\bf{H}}{\bf{F}}
+{\bf{I}})^{-1}\right],
\end{align}
is Schur-concave with respect to the diagonal elements of  $({\bf{F}}^{\rm{H}}{\bf{H}}^{\rm{H}}{\bf{R}}_{n}^{-1}{\bf{H}}{\bf{F}}
+{\bf{I}})^{-1}$. In this case, the following result  was derived \cite{Palomar03}.

\noindent{\bf Conclusion 2:} When the diagonal elements of ${\boldsymbol \Lambda}_{{\bf{w}}}$ are in decreasing order and the diagonal elements of the MSE matrix are in increasing order,
the optimal ${\bf{F}}$ for  the weighted MSE minimization problem under sum-power constraint in (\ref{OPT_Original}) has the following structure:
\begin{align}
\label{Palomar}
{\bf{F}}={\bf{V}}_{{\boldsymbol{\mathcal{H}}}}{\boldsymbol\Lambda}_{\bf{F}}.
\end{align}
With this optimal structure, the optimization problem (\ref{OPT_Original}) can be reduced to the power allocation problem and its solution was shown to be the water-filling one \footnote{Due to the space limitation, the derivation is not given and the interested readers are referred to\cite{Palomar03}.}.

\subsection{Comments on Majorization Theory Based Method}
\subsubsection{Ordering of the Weights and Diagonal Elements of the MSE Matrix}
The optimal structure in Conclusion 2 is crucial in solving the weighted MSE minimization problem. But it should be emphasized that to use the result, the conditions on the orderings of the diagonal elements of $({\bf{F}}^{\rm{H}}{\bf{H}}^{\rm{H}}{\bf{R}}_{n}^{-1}{\bf{H}}{\bf{F}}
+{\bf{I}})^{-1}$ and the reversal ordering of the diagonal elements of ${\boldsymbol \Lambda}_{{\bf{w}}}$ must be satisfied\footnote{If the orderings are violated, direct adoption of the result leads to wrong solution, as has been seen in some literature.}. For the general case when the diagonal elements of the two matrices are not ordered thus, one cannot conclude that $f(\cdot)$ is Schur-concave. One technique to conquer this drawback and solve the problem without the specific ordering conditions is to use the following matrix inequality to transfer the weighted MSE to be a Schur-concave function. \\
\textbf{Inequality 1:} If $\bf A$ and $\bf B$ are $N \times N$ positive semi-definite matrices, we have \cite{Marshall79}
\begin{align}
&  {\sum}_{i=1}^N\lambda_i({\boldsymbol A})\lambda_{N-i+1}({\boldsymbol B})  \le {\rm{Tr}}({\boldsymbol A}{\boldsymbol B}). \label{inequ_1}
\end{align}
With the  eigenvalue decompositions:
${\boldsymbol A}={\bf{U}}_{{\boldsymbol {A}}}{\boldsymbol \Lambda}_{{\boldsymbol {A}}}{\bf{U}}_{{\boldsymbol {A}}}^{\rm{H}} $ where $ {\boldsymbol \Lambda}_{{\boldsymbol {A}}} \searrow$ and
${\boldsymbol B}={\bf{\tilde U}}_{{\boldsymbol { B}}}{\boldsymbol {\tilde \Lambda}}_{{\boldsymbol {B}}}{\bf{\tilde U}}_{{\boldsymbol {B}}}^{\rm{H}}$  where  ${\boldsymbol{\tilde \Lambda}}_{{\boldsymbol {B}}} \nearrow$, the equality in (\ref{inequ_1}) holds when ${\bf{U}}_{{\boldsymbol {A}}}=\bf{\tilde U}_{{\boldsymbol {B}}}^{\rm{H}}$.

Based on Inequality 1, the weighted MSE can be transferred into a Schur-concave function for arbitrary orderings of the diagonal elements of $({\bf{F}}^{\rm{H}}{\bf{H}}^{\rm{H}}{\bf{R}}_{n}^{-1}{\bf{H}}{\bf{F}}
+{\bf{I}})^{-1}$ and ${\boldsymbol \Lambda}_{{\bf{w}}}$ via a permutation. By following the framework proposed  in \cite{Palomar03} for Schur-concave functions, the following result can be drawn.

\noindent  \textbf{Comment 3:} The weighted MSE can be understood as a Shur-concave function of the diagonal elements of the MSE matrix without a specific ordering. The optimal ${\bf{F}}$ for  the weighted MSE minimization problem in (\ref{OPT_Original}) has the following structure
\begin{align}
\label{general-structuer}
{\bf{F}}={\bf{V}}_{{\boldsymbol{\mathcal{H}}}}{\boldsymbol\Lambda}_{\bf{F}}{\bf{U}}_{\rm{Per}}^{\rm{H}},
\end{align}where ${\bf{U}}_{\rm{Per}}$ is a permutation matrix determined by the order of the diagonal elements of ${\boldsymbol \Lambda}_{\bf{w}}$.

It is obvious that when the diagonal elements of ${\boldsymbol \Lambda}_{{\bf{w}}}$ are in decreasing order and the singular values of the SVD in (\ref{channel_decomp}) are in decreasing order, we have ${\bf{U}}_{\rm{Per}}={\bf{I}}$. In general cases, further derivation of ${\bf{U}}_{\rm{Per}}$ is necessary.

With the optimal structure in (\ref{general-structuer}),  the next step is to use it in the weighted MSE matrix and solve the power allocation problem, i.e., finding the diagonal elements of ${\boldsymbol\Lambda}_{\bf{F}}$. It should be emphasized that  along with the optimal structure in Conclusion 2 or Comment 3, there is an implicit constraint that when the diagonal elements of ${\boldsymbol \Lambda}_{{\bf{w}}}$ are in decreasing order the diagonal elements of ${\bf{F}}^{\rm{H}}{\bf{H}}^{\rm{H}}{\bf{R}}_{n}^{-1}{\bf{H}}{\bf{F}}$ must be in decreasing order. This constraint needs to be taken into account in the power allocation process. Thus, after using the optimal structure, the optimization problem (\ref{OPT_Original}) becomes to be the following one
\begin{align}
\label{scalar_optimization}
& \min_{f_i^2} \ \  \sum_{i=1}^N\frac{w_i}{1+f_i^2h_i^2} \nonumber\\
& \ {\rm{s.t.}} \ \ {\rm{diag}}\{[f_1^2h_1^2,\cdots,f_N^2h_N^2]^{\rm{T}}\} \searrow \nonumber \\
&\ \ \ \ \ \ \  \sum_i f_i^2\le P,
\end{align}
where ${\rm{diag}}\{[f_1^2h_1^2,\cdots,f_N^2h_N^2]^{\rm{T}}\} \searrow $ is the implicit constraint.

In general, the ordering constraint is challenging for the optimization. One way is to relax the constraint first and compute the optimal solution without it. If the solution automatically satisfies the constraint, it is thus the optimal solution. Based on water-filling, the solution of (\ref{scalar_optimization}) after the relaxation satisfies
\begin{align}
\label{weighted_mse_waterfilling}
f_i^2h_i^2=\left(\sqrt{\frac{w_ih_i^2}{\mu }}-1\right)^{+}.
\end{align}For one  special case considered in \cite{Palomar03}, where $w_i$'s and $h_i$'s are both in decreasing order, $f_i^2h_i^2$'s are obviously in decreasing order and thus the ordering constraint is automatically satisfied. Actually, for all weighted MSE optimization problems with the optimal SVD structure in Comment 3 the authors' have encountered, the implicit condition is automatically satisfied by the water-filling solution via numerical check\cite{Tang07,Guan08}. But this cannot be proved to be a general result. For example, for weighted MSE optimization under more complicated system models, e.g., multi-hop MIMO networks and networks with channel errors, it is unclear whether the implicit condition is automatically satisfied due to the complicated nature of the water-filling solution. Thus, for the completeness of the mathematical derivation of the optimal solution,  a check on the constraint ${\rm{diag}}\{[f_1^2h_1^2,\cdots,f_N^2h_N^2]^{\rm{T}}\} \searrow $ is necessary. \\
{\bf Comment 4:} In using the optimal structure in (\ref{general-structuer}) to solve the weighted MSE optimization problem, for the obtained the water-filling power allocation result to be optimal, it should satisfy the implicit ordering condition that is used in the derivation of (\ref{general-structuer}).

%

\subsubsection{On the Applicability of Majorization Theory Based Method}
Another issue with the optimal structure in (\ref{general-structuer}) is its limited applications in the variations and generalizations of the weighted MSE minimization. To achieve (\ref{general-structuer}), the MSE matrix formulation and power constraint must be constrained to be the ones given in (\ref{OPT_Original}). Otherwise, the result may be misused. In the following, two examples are shown to support this.

%

\noindent \textbf{Case 3: Weighted MSE Minimization under Per-Antenna Constraint.}

Consider the weighted MSE minimization under per-antenna power constraints in the following form:
\begin{align}
& \min_{{\bf{F}}} \ \ {\rm{Tr}}\left[{\boldsymbol \Lambda}_{{\bf{w}}}({\bf{F}}^{\rm{H}}{\bf{H}}^{\rm{H}}{\bf{R}}_{n}^{-1}{\bf{H}}{\bf{F}}
+{\bf{I}})^{-1}\right]\nonumber \\
& \ \  {\rm{s.t.}} \ \ [{\bf{F}}{\bf{F}}^{\rm{H}}]_{n,n}\le P_n.
\end{align} The optimal solution has the following structure \cite{XingTSP201503}
\begin{align}
{\bf{F}}={\boldsymbol{\Lambda}}_{\rm{P}}^{-1/2}{\bf{V}}_{{\boldsymbol{\mathcal{\tilde H}}}}{\boldsymbol\Lambda}_{\bf{F}}{\bf{U}}_{\rm{Per}}^{\rm{H}},
\end{align}where ${\boldsymbol{\Lambda}}_{\rm{P}}^{-1/2}$ is a diagonal matrix with positive diagonal elements and the unitary matrix ${\bf{V}}_{{\boldsymbol{\mathcal{\tilde H}}}}$ is defined based on the following SVD
\begin{align}
{\bf{R}}_{\bf{n}}^{-1/2}{\bf{H}}{\boldsymbol{\Lambda}}_{\rm{P}}^{-1/2}=
{\bf{U}}_{{\boldsymbol{\mathcal{\tilde H}}}}{\boldsymbol \Lambda}_{{\boldsymbol{\mathcal{\tilde H}}}}{\bf{V}}_{{\boldsymbol{\mathcal{\tilde H}}}}
^{\rm{H}} \ \ {\rm{with}} \ \ {\boldsymbol \Lambda}_{{\boldsymbol{\mathcal{\tilde H}}}} \searrow.
\end{align} It can be concluded that although based on Inequality 1, the objective function can be transferred to be Schur-concave and the objective function has the same formulation as that in (\ref{OPT_Original}), Conclusion 2 or Comment 3 does not hold due to the per-antenna power constraint.

\noindent \textbf{Case 4: Robust MIMO transceiver optimization}

With imperfect CSI, channel estimation errors may seriously degrade system performance and robust designs play an important role.
When the columns of the channel estimation error matrix are correlated, the weighted MSE minimization problem can be formulated as follows \cite{XingTSP201501}
\begin{align}
& \min_{{\bf{F}}} \ \ {\rm{Tr}}\left[{\boldsymbol \Lambda}_{{\bf{w}}}\left(\frac{{\bf{F}}^{\rm{H}}{\bf{H}}^{\rm{H}}{\bf{H}}{\bf{F}}}
{\sigma_n^2
+{\rm{Tr}}({\bf{F}}{\bf{F}}^{\rm{H}}{\boldsymbol \Phi})}
+{\bf{I}}\right)^{-1}\right]\nonumber \\
& \ \  {\rm{s.t.}} \ \ {\rm{Tr}}({\bf{F}}{\bf{F}}^{\rm{H}}) \le P,
\end{align}where ${\boldsymbol \Phi}$ is the column-correlation matrix of the  channel error matrix. The optimal solution has the following structure \cite{XingTSP2015,XingTSP2013}:
\begin{align}\
\label{45}
{\bf{F}}=({\boldsymbol \Phi}P+\sigma_n^2{\bf{I}})^{-1/2}{\bf{V}}_{{\boldsymbol{\mathcal{\hat H}}}}{\boldsymbol\Lambda}_{\bf{F}}{\bf{U}}_{\rm{Per}}^{\rm{H}}
\end{align}where  ${\bf{V}}_{{\boldsymbol{\mathcal{\hat H}}}}$ is defined based on the following SVD
\begin{align}
{\bf{H}}({\boldsymbol \Phi}P+\sigma_n^2{\bf{I}})^{-1/2}=
{\bf{U}}_{{\boldsymbol{\mathcal{\hat H}}}}{\boldsymbol \Lambda}_{{\boldsymbol{\mathcal{\hat H}}}}{\bf{V}}_{{\boldsymbol{\mathcal{\hat H}}}}
^{\rm{H}} \ \ {\rm{with}} \ \ {\boldsymbol \Lambda}_{{\boldsymbol{\mathcal{\hat H}}}} \searrow
\end{align}

Similarly, in this case, the objective function can also be transferred into a Schur-concave function and the power constraint is the same as that in (\ref{OPT_Original}). However the optimal structure is significantly different from that in Conclusion 2.

{\bf Comment 5: } Conclusion 2 and Comment 3 only apply for specific objective structure and  power constraint. The Schur-concavity of the objective function does not automatically lead to the optimal structure in them.
\subsection{Summary}

The majorization theory based method is a powerful mathematical tool. When applicable, it can result in simple derivations  of the optimal solution without the permutation ambiguity issue. However, its application has strict limitations in the specific formulation of the objective function and the power constraint, and  whether the objective is Schur-concave cannot guarantee Conclusion 2 or comment to hold. At most times, majorization theory seems to be a supplementary theory instead of decisive theory for MIMO transceiver designs.


\section{Matrix-Field Weighted MSE Model}

While traditional weighted MSE model has been widely used and leads to successful MIMO transceiver solutions, it cannot be applied for some important design objectives such  as the min-max problem, where the largest MSE of the transmit signals needs to be minimized. In the optimal structure shown in (\ref{F-decomp}) and (\ref{general-structuer}), the unitary matrix ${\bf{U}}_{\rm{per}}$ causes the largest  weight factor always being allocated to the eigenchannel with the highest quality. To achieve the min-max design, the largest weight factor should be allocated to the eigenchannel with the the lowest quality, which is impossible with the traditional weighted MSE modeling. In addition, it cannot be used for capacity-maximization design for complicated MIMO networks. Even for point-to-point MIMO systems, its use for capacity-maximization is artificial complicated. For example, the choice of the diagonal weighting matrix cannot be determined priorly with respect to the optimization. In some work, the weighting matrix is determined by just comparing the optimal solutions for capacity maximization and weighted MSE minimization. In this section, following the ideas in \cite{XingComletter}, we propose a generalized matrix-field weighted MSE model, which are free of the aforementioned limitations. Possible methods to solve problems with the generalized modeling are discussed, along with its wide applications in MIMO\ transceiver
designs. 

\subsection{Proposed Generalized Matrix-Field Weighted MSE Model}

In the traditional weighted MSE model given in (\ref{OPT_Original_1}), the off-diagonal entries of the MSE matrix have not been taken into account in the objective  function. Thus, the information contained in the off-diagonal entries is not exploited in the MIMO\ transceiver design. In overcoming this drawback, a new model named matrix-field weighted MSE is proposed recently \cite{XingComletter}, in which the weighting operation is defined based on positive semidefinite cone, in other words, the weighting operation is extended from nonnegative vectors to positive semi-definite matrices.

For the new model to be sensible, the following properties are desired.
\begin{enumerate}
\item The proposed weighted MSE matrix must be positive semi-definite, as this is the most fundamental  property of a covariance matrix.

\item The new model must include the traditional one as its special case.

\item  The model should have tractable mathematical structure. Thus linear operation with respect to the MSE\ matrix, ${\boldsymbol \Phi}_{\rm{MSE}}$, is the most appealing.
\end{enumerate}
Based on the above properties, we propose the generalized matrix-field weighted MSE matrix as follows: \begin{align}
\label{New_Weight}
{\boldsymbol \Psi}({\bf{G}},{\bf{F}})={\sum}_{k=1}^K{\bf{W}}_k^{\rm{H}}{\boldsymbol \Phi}_{\rm{MSE}}({\bf{G}},{\bf{F}})
{\bf{W}}_k+{\boldsymbol{\Xi}},
\end{align}where ${\bf{W}}_k$'s are complex matrices and ${\boldsymbol{\Xi}}$ is a Hermitian matrix. It is obvious that (\ref{New_Weight}) enjoys all the previously listed properties. To distinguish from the traditional weighted MSE modeling in (\ref{Matrix_MSE}), ${\boldsymbol \Psi}({\bf{G}},{\bf{F}})$ is referred to  as the \textit{matrix-field  weighted MSE matrix}. Compared to \cite{XingComletter}, the generalized weighted MSE model proposed here is different from two perspectives : 1) the Hermitian matrix ${\boldsymbol{\Xi}}$ is not limited to a positive semidefinite matrix, 2) the matrices ${\bf{W}}_k$'s and ${\boldsymbol{\Xi}}$ are not limited to constant matrices but can be variables to be optimized. This allows the model to cover complicated scenarios, for example, nonlinear transceiver designs and capacity maximization problems. The model  in \cite{XingComletter} is a special case of (\ref{Opt_0_0}) in which ${\bf{W}}_k$'s and ${\boldsymbol{\Xi}}$ are constant, and ${\boldsymbol{\Xi}}$ is positive semi-definite\footnote{The model in \cite{XingComletter} applies to multihop MIMO networks. Here only single-hop point-to-point MIMO systems are considered.}.

Based on (\ref{New_Weight}), the weighted MSE minimization problem is formulated as follows:
\begin{align}
\label{Opt_0_0_0}
&\min_{{\bf{G}},{\bf{F}},{\bf{W}}_k,{\boldsymbol{\Xi}}} \ \ \ {f}_{M}[{\boldsymbol \Psi}({\bf{G}},{\bf{F}})] \nonumber모\\
& \ \ \ \ \ \ {\rm{s.t.}} \ \ \ \ g_k({\bf{F}})\le P_k, \ \ k=1,\cdots,K,
\end{align}where  $f_M(\cdot)$ is an increasing matrix-monotone function, i.e., ${\boldsymbol A} \preceq {\boldsymbol B}$ $\rightarrow$ $f_M({\boldsymbol A}) \le f_M({\boldsymbol B})$ \cite{Marshall79}. When ${\bf{W}}_k$'s and ${\boldsymbol{\Xi}}$ are independent of ${\bf{G}}$, the optimal equalizer is still the LMMSE equalizer and thus the previous optimization problem can be simplified as
\begin{align}
\label{Opt_0_0}
&\min_{{\bf{F}},{\bf{W}}_k,{\boldsymbol{\Xi}}} \ \ \ {f}_{M}[{\boldsymbol \Psi}({\bf{G}}_{\rm{LMMSE}},{\bf{F}})] \nonumber모\\
& \ \ \ \  {\rm{s.t.}} \ \ \ \ g_k({\bf{F}})\le P_k, \ \ k=1,\cdots,K,
\end{align}
where
\[
{\boldsymbol \Psi}({\bf{G}}_{\rm{LMMSE}},{\bf{F}})={\sum}_{k=1}^K{\bf{W}}_k^{\rm{H}}(
{\bf{F}}^{\rm{H}}{\bf{H}}^{\rm{H}}{\bf{R}}_n^{-1}{\bf{H}}{\bf{F}}+{\bf{I}})^{-1}{\bf{W}}_k+{\boldsymbol \Xi}.
\]
\subsection{Solution to the Generalized Matrix-Field Weighted MSE Problem}
In \cite{XingComletter}, the majorization theory based method is used for the matrix-field weighted MSE minimization. In this subsection, we discuss how to solve the generalized problem in (\ref{Opt_0_0}).

For a general objective function $f_M(\cdot)$, by introducing an auxiliary unitary matrix ${\bf{Q}}$, where ${\bf{F}}={\bf{X}}{\bf{Q}}^{\rm{H}}$, the matrix-field weighted MSE is reformulated as
\begin{align}
&{\boldsymbol \Psi}({\bf{G}}_{\rm{LMMSE}},{\bf{F}})
={\sum}_{k=1}^K{\bf{W}}_k^{\rm{H}}{\bf{Q}}({\rm{diag}}\{{\boldsymbol \lambda}({\bf{X}}^{\rm{H}}{\bf{H}}^{\rm{H}}
{\bf{R}}_{n}^{-1}
{\bf{H}}{\bf{X}})\}+{\bf{I}})^{-1}{\bf{Q}}^{\rm{H}}{\bf{W}}_k+{\boldsymbol \Xi}.
\end{align}For the convenience of presentation, we focus on a simple case with sum-power constraint. The optimization problem (\ref{Opt_0_0}) is rewritten as
\begin{align}
\label{Opt_0_1}
&\min_{{\bf{X}},{\bf{Q}},{\bf{W}}_k,{\boldsymbol \Xi}} \ \ \ {f}_{M}[{\sum}_{k=1}^K{\bf{W}}_k^{\rm{H}}{\bf{Q}}({\rm{diag}}\{{\boldsymbol \lambda}({\bf{X}}^{\rm{H}}{\bf{H}}^{\rm{H}}
{\bf{R}}_{n}^{-1}
{\bf{H}}{\bf{X}})\}+{\bf{I}})^{-1}{\bf{Q}}^{\rm{H}}{\bf{W}}_k+{\boldsymbol \Xi}] \nonumber모\\
& \ \ \ \  {\rm{s.t.}} \ \ \ \ {\rm{Tr}}({\bf{X}}{\bf{X}}^{\rm{H}}) \le P.
\end{align}

The optimal solution of ${\bf{Q}}$ is determined by the specific objective functions \cite{XingTSP201501}. It has been shown that for any given unitary matrix ${\bf{Q}}$,  the optimal ${\bf{X}}$ for (\ref{Opt_0_1}) is one of the Pareto optimal solutions of  the following matrix-monotonic optimization problem \cite{XingTSP201501}
\begin{align}
\label{matrix_monotonic}
& \max_{{\bf{X}}} \ \ {\bf{X}}^{\rm{H}}{\bf{H}}^{\rm{H}}{\bf{R}}_{n}^{-1}
{\bf{H}}{\bf{X}}\nonumber \\
& \ {\rm{s.t.}} \ \ \ {\rm{Tr}}({\bf{X}}{\bf{X}}^{\rm{H}}) \le P.
\end{align}This is a multi-objective optimization problem and the maximum of the objective value is defined in the semi-definite matrix cone \cite{XingTSP201501}. The matrix ${\bf{X}}^{\rm{H}}{\bf{H}}^{\rm{H}}{\bf{R}}_{n}^{-1}
{\bf{H}}{\bf{X}}$ can be interpreted as a matrix version of the signal-to-noise-ration (SNR) \cite{XingTSP201503}.
Since the power constraint is unitary invariant, there is no constraint on the EVD unitary matrix of ${\bf{X}}^{\rm{H}}{\bf{H}}^{\rm{H}}{\bf{R}}_{n}^{-1}
{\bf{H}}{\bf{X}}$. In other words, maximizing ${\bf{X}}^{\rm{H}}{\bf{H}}^{\rm{H}}{\bf{R}}_{n}^{-1}
{\bf{H}}{\bf{X}}$ is equivalent to maximizing ${\boldsymbol \lambda}({\bf{X}}^{\rm{H}}{\bf{H}}^{\rm{H}}{\bf{R}}_{n}^{-1}
{\bf{H}}{\bf{X}})$. The following result on Pareto optimal solutions of (\ref{matrix_monotonic}) has been proved  \cite{XingTSP201501}.

\noindent {{\textbf{Conclusion 4:}}} Any Pareto optimal solution of (\ref{matrix_monotonic}) satisfies the following structure \cite{XingTSP201501}:
\begin{align}
{\bf{X}}={\bf{V}}_{{\boldsymbol{\mathcal{H}}}}
{\boldsymbol{\Lambda}}_{\bf{X}}{\bf{U}}_{\rm {Arb}}^{\rm{H}},
\end{align}
where ${\bf{V}}_{{\boldsymbol{\mathcal{H}}}}$ is defined in (\ref{channel_decomp}), ${\bf{U}}_{\rm {Arb}}$ is an arbitrary unitary matrix, and the diagonal matrix ${\boldsymbol{\Lambda}}_{\bf{X}}$ satisfies
\begin{align}
{\boldsymbol{\Lambda}}_{\bf{X}}^{\rm{T}}{\boldsymbol{\Lambda}}
_{{\boldsymbol{\mathcal{H}}}}^{\rm{T}}{\boldsymbol{\Lambda}}
_{{\boldsymbol{\mathcal{H}}}}{\boldsymbol{\Lambda}}_{\bf{X}} \searrow.
\end{align}

We would like to highlight that for the general case where in   (\ref{channel_decomp}) the diagonal elements of ${\boldsymbol \Lambda}_{\boldsymbol{\mathcal{H}}}$ are not in decreasing order, the diagonal elements of ${\boldsymbol{\Lambda}}_{\bf{X}}^{\rm{T}}{\boldsymbol{\Lambda}}
_{{\boldsymbol{\mathcal{H}}}}^{\rm{T}}{\boldsymbol{\Lambda}}
_{{\boldsymbol{\mathcal{H}}}}{\boldsymbol{\Lambda}}_{\bf{X}}$ should have the same order as that of ${\boldsymbol \Lambda}_{\boldsymbol{\mathcal{H}}}$. Since ${\bf{U}}_{\rm {Arb}}$ is an arbitrary unitary matrix, we can simply set ${\bf{U}}_{\rm {Arb}}={\bf{I}}$.

After finding the optimal SVD structure, the diagonal elements of ${\boldsymbol{\Lambda}}_{\bf{X}}$ denoted as $x_i$'s can be solved based on multi-objective optimization theory. Let $\tilde{w}_i$'s be the nonnegative weighting factors based on which Pareto optimal solutions can be computed using scalarization. For any set of $\tilde{w}_i$'s satisfying $\ {\tilde w}_1h_1^2\ge \cdots \ge {\tilde w}_Nh_N^2$, the corresponding Pareto optimal solution is\cite{XingTSP201501}
\begin{align}
\label{WF-solution-a}
x_i^2=\left(\sqrt{\frac{{\tilde w}_i}{\mu h_i^2}}-\frac{1}{h_i^2}\right)^+.
\end{align}
The proof is provided in Appendix \ref{app-C}. Note that as discussed in Appendix~\ref{app-C}, different from the Pareto optimal solution of traditional weighted MSE model given in (\ref{WF-solution}) the ${\tilde w}_i$'s in (\ref{WF-solution-a}) are required to make ${\tilde w}_ih_i^2$'s in decreasing order instead of themselves being in decreasing order.
 This means that the Pareto optimal solution set of the traditional weighted MSE model is a subset of that of the matrix-field weighted MSE model.

\subsubsection{Weighted MSE Minimization} When ${\bf{W}}_k$'s and ${\boldsymbol \Xi}$ are constants and ${f}_M[{\boldsymbol \Psi}({\bf{G}}_{\rm{LMMSE}},{\bf{F}})]={\rm{Tr}}[{\boldsymbol \Psi}({\bf{G}}_{\rm{LMMSE}},{\bf{F}})]$, the problem in (\ref{Opt_0_0}) becomes the weighted MSE minimization problem (where the weight matrix is not necessarily diagonal) as follows:
\begin{align}
\label{Opt_1}
& {\min_{\bf{F}}} \ \ \ {\rm{Tr}}[{\bf{W}}(
{\bf{F}}^{\rm{H}}{\bf{H}}^{\rm{H}}{\bf{R}}_n^{-1}{\bf{H}}{\bf{F}}+{\bf{I}})^{-1} ] \nonumber \\
& \ {\rm{s.t.}} \ \ \ {\rm{Tr}}({\bf{F}}{\bf{F}}^{\rm{H}}) \le P,
\end{align}
where ${\bf{W}}=\sum_{k=1}^K{\bf{W}}_k{\bf{W}}_k^{\rm{H}}$. Given the following EVD of the weighting matrix ${\bf{W}}$
\begin{align}
{\bf{W}}={\bf{U}}_{\bf{W}}{\boldsymbol \Lambda}_{\bf{W}}{\bf{U}}_{\bf{W}}^{\rm{H}} \ \
\text{with} \ \ {\boldsymbol \Lambda}_{\bf{W}} \searrow,
\end{align}with the help of Inequality 1, the optimal ${\bf{Q}}$ can be derived. Together with Conclusion 4, the following result can be obtained.

\noindent {\textbf{Conclusion 5:}} The optimal solution of (\ref{Opt_1}) has the following SVD structure: \begin{align}
{\bf{F}}={\bf{V}}_{{\boldsymbol{\mathcal{H}}}}
{\boldsymbol\Lambda}_{\bf{F}}{\bf{U}}_{\bf{W}}^{\rm{H}}.
\end{align}

Different from Conclusions 1 and 2,  the unitary matrix ${\bf{U}}_{\bf{W}}$ in Conclusion 5 can be an arbitrary unitary matrix and is not necessarily a permutation matrix. It depends on the weight matrix ${\bf{W}}$.




\subsection{Applications of the Generalized Matrix-Field Weighted MSE Model}
The proposed generalized matrix-field weighted MSE model introduces more degrees of freedom to the MIMO transceiver design than the traditional model, and enlarges the application range. As has been shown in the previous subsection, it can naturally cover the weighted MSE minimization problem. In what follows, its other applications that cannot be covered by the traditional model are explained.
\subsubsection{Min-Max problem} For min-max optimization, the objective is a Schur-convex function of the diagonal elements of the MSE matrix and the optimal solution makes the the diagonal elements of the MSE matrix  all equal \cite{Palomar03}. In other words, the min-max optimization is equivalent to minimizing the sum of the diagonal elements of the MSE matrix and meanwhile keeping the diagonal elements equal\cite{Palomar03}. It can be realized with the proposed matrix-field model. By setting ${\bf{U}}_{\bf{W}}={\bf{U}}_{\rm{DFT}}$ where ${\bf{U}}_{\rm{DFT}}$ is the discrete Fourier transform (DFT) matrix and having the diagonal elements of ${\boldsymbol \Lambda}_{\bf{W}}$ the same, the weighted MSE objective in (\ref{Opt_1}) is equivalent to sum MSE, and the diagonal elements of resultant MSE matrix are equal because of ${\bf{U}}_{\rm{DFT}}$.

\subsubsection{Nonlinear Transceiver Optimization}

As pointed out in \cite{XingTSP2015}, nonlinear transceiver designs with THP or DFE can be realized by matrix-field weighted MSE model. With the following setting: $K=1$, ${\bf{W}}_1={\bf{C}}$ and ${\boldsymbol{\Xi}}={\bf{0}}$, the matrix-field weight MSE matrix can be written as
\begin{align}
{\boldsymbol \Psi}({\bf{G}}_{\rm{LMMSE}},{\bf{F}})={\bf{C}}({\bf{F}}^{\rm{H}}{\bf{H}}^{\rm{H}}{\bf{R}}_{n}^{-1}{\bf{H}}{\bf{F}}
+{\bf{I}})^{-1}{\bf{C}}^{\rm{H}},
\end{align}where ${\bf{C}}$ is a lower triangular matrix.  By peoperly defining $f_M(\cdot)$ to be a function of the diagonal entries of the matrix variable,  the nonlinear transceiver designs with THP or DFE are obtained from (\ref{Opt_0_1}), which is given as follows:
\begin{align}
&모\min_{{\bf{F}},{\bf{C}}} \ \ f({\bf{d}} [({\bf{C}}({\bf{F}}^{\rm{H}}{\bf{H}}^{\rm{H}}{\bf{R}}_{n}^{-1}{\bf{H}}{\bf{F}}
+{\bf{I}})^{-1}{\bf{C}}^{\rm{H}}]) \nonumber \\
& \ \ {\rm{s.t.}} \ \ {\rm{Tr}}({\bf{F}}{\bf{F}}^{\rm{H}})\le P,
\end{align}where $f(\cdot)$ is a multiplicatively Schur-convex or multiplicatively Schur-concave function \cite{XingTSP2015}.

\subsubsection{Capacity Maximization}
\label{sec-extension}
Capacity is an important and widely used performance metric for transceiver optimization. With the proposed generalized model, it can be realized via the matrix-field weighted MSE minimization.
When $K=1$,  ${\boldsymbol{\Xi}}=-\frac{1}{N}{\rm{log}}|{\bf{W}}_1{\bf{W}}_1^{\rm{H}}|{\bf{I}}$, and
\begin{align}
 {f}_{M}[{\boldsymbol \Psi}({\bf{G}},{\bf{F}})]={\rm{Tr}}[{\boldsymbol \Psi}({\bf{G}},{\bf{F}})]={\rm{Tr}}({\bf{W}}_1^{\rm{H}}{\boldsymbol \Phi}_{\rm{MSE}}({\bf{G}},{\bf{F}})
{\bf{W}}_1)-{\rm{log}}|{\bf{W}}_1{\bf{W}}_1^{\rm{H}}|,
\end{align}
the generalized matrix-field weighted MSE minimization problem becomes
\begin{align}
\label{MSE_Capacity}
&\min_{{\bf{G}},{\bf{F}},{\bf{W}}_1} \ \ \ {\rm{Tr}}({\bf{W}}_1^{\rm{H}}{\boldsymbol \Phi}_{\rm{MSE}}({\bf{G}},{\bf{F}})
{\bf{W}}_1)-{\rm{log}}|{\bf{W}}_1{\bf{W}}_1^{\rm{H}}| \nonumber모\\
& \ \ {\rm{s.t.}} \ \ \ \ \ \ {\rm{Tr}}({\bf{F}}{\bf{F}}^{\rm{H}})\le P.
\end{align}
This has been proven to be equivalent to the capacity maximization \cite{Shi2011,Christensen2008}. It is worth highlighting that it is also possible to exploit the logic in (\ref{MSE_Capacity}) to realize  capacity maximization for general MIMO networks consisting of multiple source nodes, multiple relay nodes, and multiple destination nodes with multihop AF transmissions under imperfect CSI \cite{XingMiletary2010,XingIET}. By introducing multiple precoding matrices for the transmission of the multiple hops in the model, the jointly transceiver matrix design of the MIMO relay network can be formulated. In \cite{XingMiletary2010,XingIET}, an iterative algorithm has been proposed for the case where $f_M(\cdot)$ is the trace function and ${\bf{W}}_m$'s are fixed.


\subsubsection{Other Applications and Extensions}
 Other than the aforementioned applications, the ARQ based MIMO transceiver optimization  can also be understood as matrix-field weighting operations \cite{XingComletter,Sun2007}.

\subsection{Summary}
The proposed generalized matrix-field weighted MSE model can help cover more MIMO transceiver design problems with a variety of objective functions and  complicated system scenarios, such as capacity-maximization and nonlinear transceiver operations. The model also allows tractable solutions. With the help of majorization theory and other optimization tools, the optimal structure of the precoding matrix can be derived. But the determination of the permutation matrix is highly involved and needs to be scrutinized for each specific scenario.

\section{Numerical Results and Evaluations}

There are different opinions on the effectiveness of the weighted MSE model for MIMO transceiver designs. Different from the capacity and the BER criteria, MSE is connected to system performance less directly. In existing work, many papers advocate that for linear MIMO transceiver designs, the weighted MSE minimization can effectively realize the BER minimization. In this section, simulation results on point-to-point MIMO\ systems are given to show that the performance advantage of the weighted MSE model is usually based on proper choice of the weighting factors. We also use simulation results to justify some of the analytical results in previous sections. In all simulations, the weighted MSE\ minimization problem is considered under a total power constraint $P$ as formulated in (\ref{OPT_Original}). The SNR is defined as $P/\sigma_n^2$.
\begin{figure}[!t]
\centering
\includegraphics[width=0.6\textwidth]{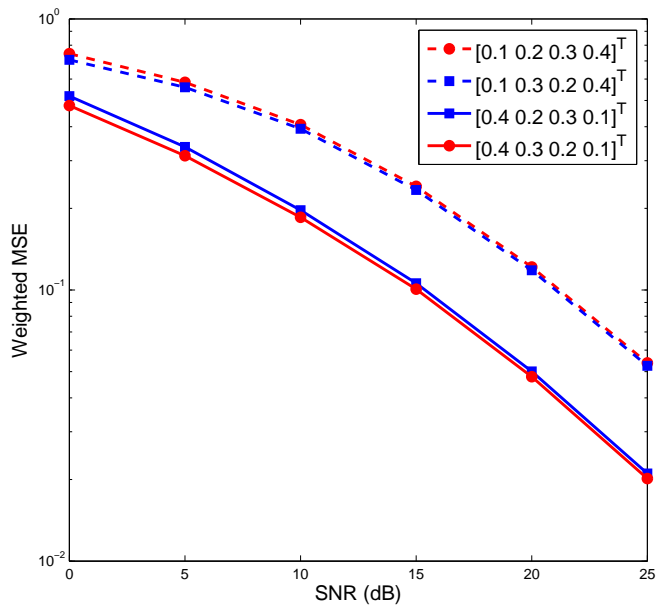} \\
\caption{The weight MSEs of a $4\times 4$ MIMO system. Lagrange multiplier method is used with 4 different permutation matrix to show the permutation ambitious effect.}
\label{Fig_1}
\end{figure}

Firstly, a $4 \times 4$ MIMO system is simulated to show the permutation ambiguous effect. Four data streams are transmitted each with the QPSK modulation. The weights for the MSEs of the four data streams are set as $0.1, 0.2,0.3,0.4$.
The Lagrange multiplier method is applied for the optimization, where Conclusion 1 is used for the optimal SVD structure of the precoding matrix. Four different permutation matrices ${\bf{U}}_{\rm{Per}}$ are tried. For each ${\bf{U}}_{\rm{Per}}$, the optimal power allocation is calculated and the resulted weighted MSE is calculated and shown in Fig.~\ref{Fig_1}. Notice that the actually effect of the unitary matrix ${\bf{U}}_{\rm{Per}}$
is to realize the allocation of the weighting factors, i.e., the diagonal elements of ${\boldsymbol \Lambda}_{\bf{w}}$, over different eigenchannels. Thus to save the space, instead of the adopted ${\bf{U}}_{\rm{Per}}$ matrices, we provide their corresponding orderings of the weighting factors in the legend, while the amplitudes of the eigenchannels are in a decreasing order.  It can be clearly seen from the figure that permutation ambiguous effect significantly affects the system performance. Improper choice of permutation matrix ${\bf{U}}_{\rm{Per}}$ can result in serious performance degradation.
\begin{figure}[!t]
\centering
\includegraphics[width=0.6\textwidth]{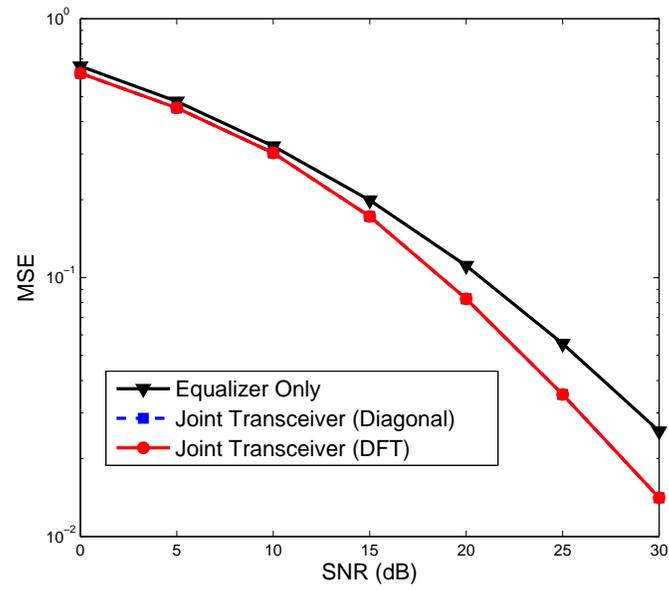} \\
\caption{The sum-MSEs of a $4\times 4$ MIMO system with different transceiver designs when the generalized matrix-field weighted MSE model is used and Conclusion 5 is adopted for the solution.}
\label{Fig_2}
\end{figure}

The next simulation is also for a $4\times 4$ MIMO system with 4 data streams and QPSK constellation. The sum-MSE is used as the objective criterion. It is on the result in Conclusion 5 for the generalized matrix-field weighted MSE model. Three solutions are simulated to make comparisons. The first one, named equalizer only, is the design in which only the LMMSE equalizer is used at the destination while the source precoder matrix is a scaled identity matrix. The other two use both the LMMSE equalizer and the optimal precoder based on Conclusion 5. Two representative unitary matrices are chosen, i.e., the identity matrix and the DFT matrix. The corresponding solutions are named joint transceiver (Diagonal) and the joint transceiver (DFT), respectively. As shown in Fig.~\ref{Fig_2}, the joint solutions have better performance than the equalizer only solution.
Furthermore, the two joint solutions have the same sum-MSE performance, verifying the conclusion that ${\bf{U}}_{\bf{w}}$ can be an arbitrary unitary matrix.

\begin{figure}[!t]
\centering
\includegraphics[width=0.6\textwidth]{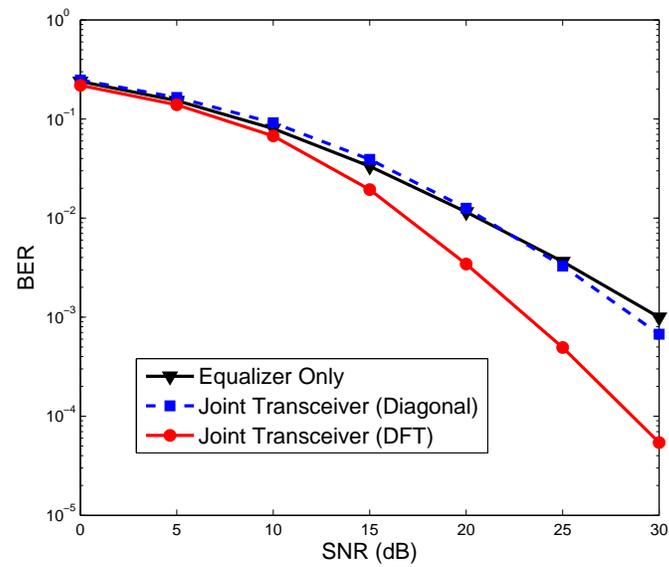} \\
\caption{The BERs of different transceiver designs for a $4 \times 4$ MIMO system.}
\label{Fig_3}
\end{figure}

\begin{figure}[!t]
\centering
\includegraphics[width=0.6\textwidth]{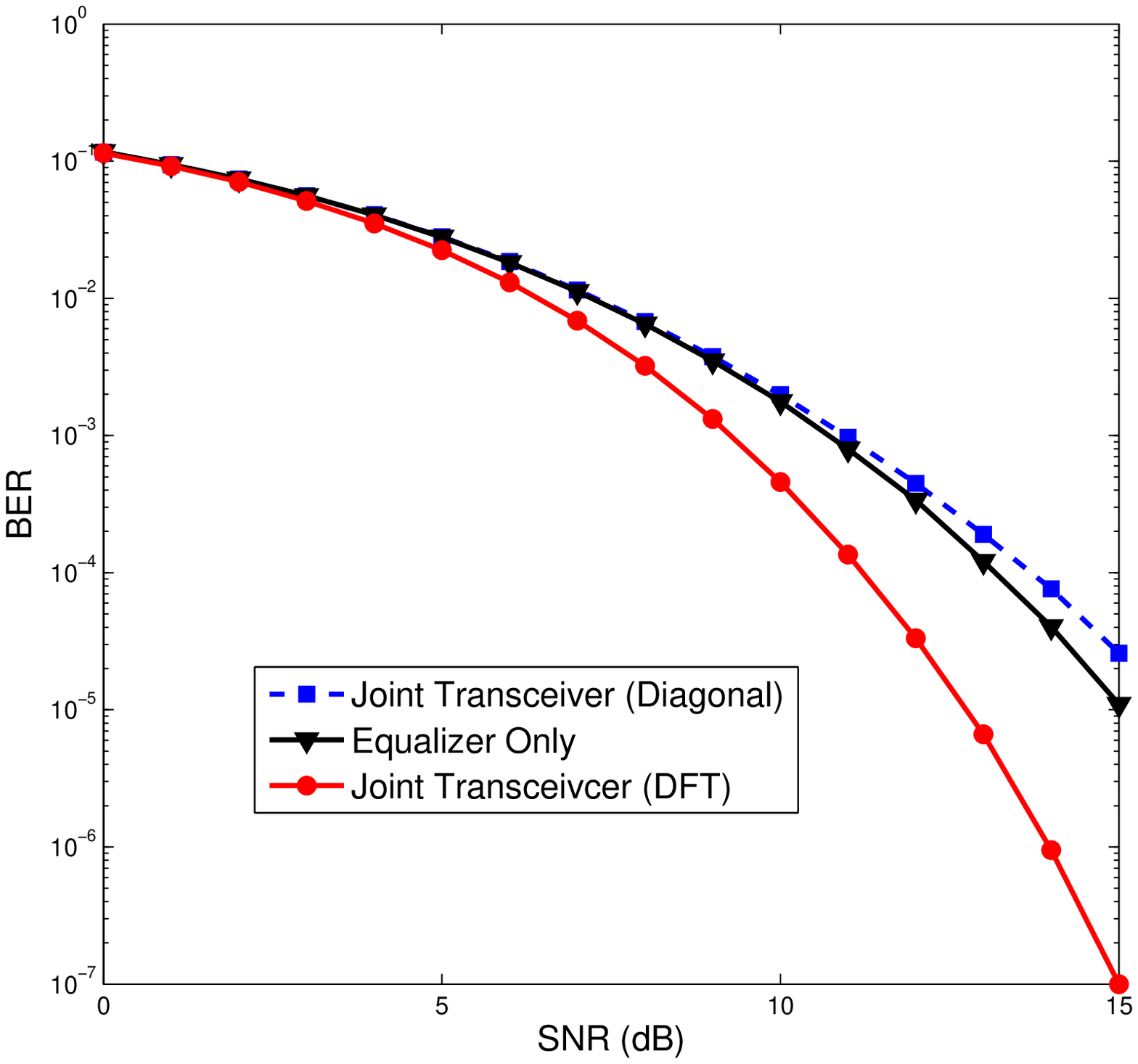} \\
\caption{The BERs of different transceiver designs for a $8 \times 4$ MIMO system.}
\label{Fig_4}
\end{figure}

In Fig.~\ref{Fig_3}, the BERs of the same system under the three solutions are shown. It is interesting to see that the equalizer only design performs better than the joint design with  ${\bf{U}}_{\bf{w}}={\bf{I}}$ at low SNR. We also simulated another $8 \times 4$ MIMO system and the BERs are shown in Fig.~\ref{Fig_4}. It can be seen that the equalizer only design performs better than the joint design with  ${\bf{U}}_{\bf{w}}={\bf{I}}$ even at high SNR. This discovery seems to say that weighted MSE is not a meaningful criterion. However, in both Figs. 3 and 4, the joint design with ${\bf{U}}_{\bf{w}}={\bf{U}}_{\rm{DFT}}$ performs significantly better than the other two algorithms. This implies two facts. First, the weighted MSE is a useful performance metric since the joint design with the DFT matrix is  one of the weighted MSE minimizing solutions. The second fact is that a careful selection of the candidate weighted-MSE minimizing solutions is necessary to achieve the high BER performance.

\section{Conclusions}
 Weighted mean-square-error (MSE) minimization is a widely used performance metric in MIMO transceiver designs that reflects how accurate signals can be recovered from noise corrupted observations. In this paper, we first reviewed the weighted MSE models and two major methods  in finding the solutions, e.g., the Lagrange multiplier method and the majorization theory based method. Then, several critical problems and facts that are often neglected in related work were pointed out. The advantages and weaknesses of the methods were analyzed. In addition, a new generalized matrix-field weighted MSE model was proposed, which covers many more applications with different objective functions and system scenarios. Possible solutions to the proposed modeling was also discussed. As MIMO technology becomes an important ingredient of more complicated upcoming communication configurations, these models, solutions, and their  limitations are important to be understood for future MIMO research.

\appendices

\section{}
\label{multi-objective optimization}
\label{app-C}
By using Conclusion 4 and after an equivalent transformation in the sense of the Pareto optimal solution set, the multi-objective optimization problem in (\ref{matrix_monotonic}) can be reduced to the following\cite{XingTSP201501}:
\begin{align}
\label{multi-prob-1}
& \min_{{{x}}_i^2} \ \ \left[\frac{1}{x_1^2h_1^2+1},\cdots,\frac{1}{x_N^2h_N^2+1}\right]^{\rm{T}}\nonumber \\
& \ {\rm{s.t.}} \ \ {\rm{diag}}\{[x_1^2h_1^2,\cdots,x_N^2h_N^2]^{\rm{T}}\} \searrow \nonumber \\
&\ \ \ \ \ \ \sum_{i=1}^N x_i^2 \le P.
\end{align} Because of the first constraint, directly computing the Pareto optimal solution set of (\ref{multi-prob-1}) is a challenging task. Now consider the following multi-objective optimization problem where the first constraint in (\ref{multi-prob-1}) is relaxed:
\begin{align}
\label{multi_objective}
& \min_{{{x}}_i^2} \ \ \left[\frac{1}{x_1^2h_1^2+1},\cdots,\frac{1}{x_N^2h_N^2+1}\right]^{\rm{T}}\nonumber \\
& \ {\rm{s.t.}} \ \ \sum_{i=1}^N x_i^2 \le P.
\end{align}
The relaxed optimization problem is much easier to deal. In the following, the Pareto optimal solution set of (\ref{multi-prob-1}) will be computed based on the Pareto optimal solution set of (\ref{multi_objective}). First, it is obvious that the Pareto optimal solution set of (\ref{multi-prob-1}) is a subset of that of (\ref{multi_objective}). More specifically, the Pareto optimal solution of (\ref{multi-prob-1}) is the Pareto optimal solution of (\ref{multi-prob-1}) satisfying  ${\rm{diag}}\{[{ x}_1^2h_1^2,\cdots,{x}_N^2h_N^2]^{\rm{T}}
\}\searrow$. 

Since each objective function is convex, the Pareto optimal solution set of (\ref{multi_objective}) can be obtained by using the following scalarization method \cite{Boyd04}
\begin{align}
& \min_{{{x}}_i^2} \ \ \sum_{i=1}^N \frac{{\tilde w}_i}{x_i^2h_i^2+1}\nonumber \\
& \ {\rm{s.t.}} \ \ \sum_{i=1}^N x_i^2 \le P
\end{align}where ${\tilde w}_i$'s are nonnegative weighting factors and  not limited to be in decreasing order \cite{Boyd04}.
By taking $x_i^2$ as variables, the optimal solution of the above optimization problem is the water-filling solution with the following form
\begin{align}
\label{water_filling_appendix}
x_i^2=\left(\sqrt{\frac{{\tilde w}_i}{\mu h_i^2}}-\frac{1}{h_i^2}\right)^+.
\end{align}Based on the previous discussions, the Pareto optimal solution of (\ref{multi-prob-1}) is the Pareto optimal solution of (\ref{multi-prob-1}) satisfying  ${\rm{diag}}\{[{ x}_1^2h_1^2,\cdots,{x}_N^2h_N^2]^{\rm{T}}
\}\searrow$.  For  $x_i^2h_i^2$'s to be in decreasing order, the weighting factors should satisfy $
{\tilde w}_1h_1^2\ge \cdots \ge {\tilde w}_Nh_N^2$. It should be highlighted that this relationship does not mean that $w_n$'s are in decreasing order, i.e.,
\begin{align}
{\tilde w}_1h_1^2\ge \cdots \ge {\tilde w}_Nh_N^2\not\rightarrow {\tilde w}_1\ge \cdots \ge {\tilde w}_N.
\end{align}
For example, when the channel parameters $\{h_i^2\}$ are in strictly decreasing order  $x_1^2h_1^2=\cdots=x_Nh_N^2$ is a Pareto optimal objective value for (\ref{multi_objective}) and in this case, ${\tilde w}_i$'s are in increasing order instead of decreasing order. By comparing (\ref{WF-solution}) and (\ref{water_filling_appendix}), it can be concluded that the Pareto optimal solution set of the traditional weighted model is a subset of that of the matrix-field MSE model. It is because the solution given by (\ref{WF-solution}) definitely satisfies (\ref{water_filling_appendix}), but the converse is not true.

\end{document}